\pdfoutput=1
\documentclass[a4paper,11pt]{article}
\usepackage[utf8]{inputenc}
\usepackage{jcappub}
\usepackage[utf8]{inputenc}
\usepackage{amsmath,amssymb}
\usepackage{graphicx}
\usepackage{multicol}
\usepackage{hyperref}
\usepackage{physics}
\usepackage{lipsum}
\usepackage{caption}
\usepackage{subcaption}
\usepackage{accents}
\usepackage{caption}
\usepackage{subcaption}
\usepackage{multirow}

\usepackage{comment}

\newcommand{\m}{m_{\text{P}}}

\usepackage{pgfplots}
\usepgfplotslibrary{fillbetween}
\usetikzlibrary{patterns}
\pgfplotsset{compat=1.8}

\newcommand{\dotE}[1]{\accentset{\circ}{#1}}
\newcommand{\ddotE}[1]{\accentset{\circ\circ}{#1}}

\DeclareRobustCommand{\dotEforfig}[1]{\accentset{\circ}{#1}}

\newcommand{\be}{\begin{equation}} 
\newcommand{\ee}{\end{equation}}
\newcommand{\bea}{\begin{equation}\begin{aligned}} 
\newcommand{\eea}{\end{aligned}\end{equation}}
\newcommand{\ba}{\begin{eqnarray}}
\newcommand{\ea}{\end{eqnarray}}

\title{\boldmath Palatini $R^2$ Quintessential Inflation}
\author[a]{Konstantinos Dimopoulos,}
\author[b]{Alexandros Karam,}
\author[a]{Samuel S\'anchez L\'opez}
\author[b]{and Eemeli Tomberg}

\emailAdd{k.dimopoulos1@lancaster.ac.uk} 
\emailAdd{alexandros.karam@kbfi.ee} 
\emailAdd{s.sanchezlopez@lancaster.ac.uk}
\emailAdd{eemeli.tomberg@kbfi.ee}

\vspace{5cm}
\affiliation[a]{Consortium for Fundamental Physics, Physics Department,\\Lancaster University, Lancaster LA1 4YB, United Kingdom.}
\affiliation[b]{Laboratory of High Energy and Computational Physics, 
National Institute of Chemical Physics and Biophysics, R{\"a}vala pst.~10, Tallinn, 10143, Estonia}

\abstract{We construct a model of quintessential inflation in Palatini $R^2$ gravity employing a scalar field with a simple exponential potential and coupled to gravity with a running non-minimal coupling. At early times, the field acts as the inflaton, while later on it becomes the current dark energy. Combining the scalar sector with an ideal fluid, we study the cosmological evolution of the model from inflation all the way to dark energy domination. We interpret the results in the Einstein frame, where a coupling emerges between the fluid and the field, feeding energy from the former to the latter during the matter-dominated era. We perform a numerical scan over the parameter space and find points that align with observations for both the inflationary CMB data and the late-time behaviour. The final dark energy density emerges from an interplay between the model parameters, without requiring the extreme fine-tuning of the cosmological constant in $\Lambda$CDM.
}

\begin{document}

\maketitle

\section{Introduction}

The content and evolution of the Universe are very well encoded and described by the concordance cosmological model, also called $\Lambda$ Cold Dark Matter ($\Lambda$CDM). Its three major components, namely, ordinary matter (including photons and neutrinos), non-relativistic or cold dark matter (CDM), and the cosmological constant $\Lambda$ are captured by $6$ independent parameters which completely specify the model. In addition, the model assumes two phases of accelerated expansion which are taking place in the very early and late Universe: Inflation and Dark Energy.

The inflationary epoch~\cite{Starobinsky:1980te, Kazanas:1980tx, Sato:1980yn, Guth:1980zm, Linde:1981mu, Albrecht:1982wi, Linde:1983gd} was originally proposed as a solution to the shortcomings of the hot Big Bang model. In a single stroke, inflation is able to generate a flat, homogeneous and isotropic Universe without any topological defects. Moreover, it provides a natural mechanism where vacuum quantum fluctuations of the gravitational and matter fields get amplified to cosmological perturbations~\cite{Starobinsky:1979ty, Mukhanov:1981xt, Hawking:1982cz, Starobinsky:1982ee, Guth:1982ec, Bardeen:1983qw}, which later became the seeds for the cosmic microwave background (CMB) primordial anisotropy and the large-scale structure of the Universe. In its simplest realization, inflation is described by a scalar field which is minimally coupled to gravity, has a canonical kinetic term and is governed by a potential whose energy density drove the (quasi-)exponential expansion. However, recent observations by the Planck~\cite{Planck:2018jri} and BICEP/Keck~\cite{BICEP:2021xfz} collaborations have essentially ruled out many simple models like monomial chaotic inflation or natural inflation and have prompted the exploration of more complicated models. The constraint on the tensor-to-scalar ratio ($r<0.036$)~\cite{BICEP:2021xfz} suggests that the inflaton potential has to be sufficiently flat at large field values. An easy way to achieve the flattening of the potential is to either couple the inflaton field non-minimally to gravity or to add a quadratic curvature term in the action. Both of these terms are generated from quantum corrections so it is natural to include them.

Modeling inflation in modified gravity is further facilitated in the Palatini formalism.
The Palatini formulation of gravity~\cite{Palatini1919, Ferraris1982} has recently gained considerable popularity as an alternative to the usual metric formulation. It treats the metric and the connection as independent variables, which means that one has to vary the action with respect to both of them. For a minimally coupled scalar field and an action linear in $R$ the two formulations result in the same equations of motion and the connection turns out to be the Levi-Civita one. However, when the field is non-minimally coupled to gravity~\cite{Bauer:2008zj, Bauer:2010bu, Tamanini:2010uq, Bauer:2010jg, Rasanen:2017ivk, Tenkanen:2017jih, Racioppi:2017spw, Markkanen:2017tun, Jarv:2017azx, Fu:2017iqg, Racioppi:2018zoy, Carrilho:2018ffi, Kozak:2018vlp, Rasanen:2018fom, Rasanen:2018ihz, Almeida:2018oid, Shimada:2018lnm, Takahashi:2018brt, Jinno:2018jei, Rubio:2019ypq, Bostan:2019uvv, Bostan:2019wsd, Tenkanen:2019xzn, Racioppi:2019jsp, Tenkanen:2020dge, Shaposhnikov:2020fdv, Borowiec:2020lfx, Jarv:2020qqm, Karam:2020rpa, McDonald:2020lpz, Langvik:2020nrs, Shaposhnikov:2020gts, Shaposhnikov:2020frq, Gialamas:2020vto, Mikura:2020qhc, Verner:2020gfa, Enckell:2020lvn, Reyimuaji:2020goi, Karam:2021wzz, Mikura:2021ldx, Kubota:2020ehu, Saez-ChillonGomez:2021byq, Mikura:2021clt} and/or quadratic or higher curvature terms are included~\cite{Olmo:2011uz, Bombacigno:2018tyw, Enckell:2018hmo, Antoniadis:2018ywb, Antoniadis:2018yfq, Tenkanen:2019jiq, Edery:2019txq, Giovannini:2019mgk, Tenkanen:2019wsd, Gialamas:2019nly, Tenkanen:2020cvw, Lloyd-Stubbs:2020pvx, Antoniadis:2020dfq, Ghilencea:2020piz, Das:2020kff, Gialamas:2020snr, Ghilencea:2020rxc, Bekov:2020dww, Dimopoulos:2020pas, Gomez:2020rnq, Karam:2021sno, Annala:2021zdt, Lykkas:2021vax, Gialamas:2021enw, AlHallak:2021hwb, Dioguardi:2021fmr, Dimopoulos:2022tvn}, significant differences arise. In the case of the non-minimal coupling, the difference can be readily seen when one transforms the Jordan frame action to the Einstein frame one. Because the Riemann tensor only depends on the connection in the Palatini formalism, this means that the Ricci scalar (which is a contraction of the metric with the Riemann tensor) transforms differently under a Weyl transformation in the two formalisms. As a result, the scalar picks up an extra coefficient in its kinetic term which is absent in the Palatini version of the theory. Therefore, the field redefinition which renders the scalar field canonical is different and the resulting Einstein frame potential is usually flatter in the Palatini formulation. Similarly, when an $\alpha R^2$ term is added to the action, the auxiliary field which is usually introduced in order to eliminate this term turns out to be non-dynamical in the Palatini formulation, in contrast to the metric version. Consequently, while the metric theory becomes two-field and therefore complicated to analyze, in the Palatini version the auxiliary field can be eliminated through its equation of motion and the resulting action is single-field, albeit modified. The main modification concerns the inflaton potential which is divided by a factor that again renders it asymptotically flat.

$\Lambda$CDM is very simple but extremely fine-tuned. This is why alternatives have been put forward, one of the most prominent of which is quintessence; the fifth element after baryons, CDM, photons and neutrinos. Similar to the inflaton, quintessence is also a scalar field \cite{Caldwell:1997ii} (for a review see \cite{Copeland:2006wr}). Therefore, unlike the cosmological constant, quintessence corresponds to a dynamic degree of freedom. Its evolving equation of state is different from that of the other constituents of the Universe content at present (baryons, CDM, neutrinos, and photons). Depending on the ratio of its kinetic and potential energy it can be either attractive or repulsive. As in slow-roll inflation, a slowly varying quintessence field can lead to the current accelerated expansion of the Universe. The energy scale of the potential energy density of quintessence must be of the order of $(10^{-12} \ \rm{GeV})^4$ today, which is more than a hundred orders of magnitude lower than the typical scale of inflation. Moreover, being dynamical, quintessence requires determination of its initial conditions, such that it attains the desired energy density at present. This is called the coincidence requirement.

One way to account for coincidence, is to connect quintessence with inflation so that the initial conditions of quintessence are determined by the inflationary attractor. Indeed, it is economic to combine the inflationary and quintessential epochs and describe them by a single scalar field in the context of a common theoretical framework. This idea has been dubbed \emph{quintessential inflation} \cite{Peebles:1998qn}.\footnote{The idea of unifying inflation and quintessence was much older \cite{Peccei:1987mm, Wetterich:1994bg}.} 
Of course, for such a model to be consistent, the scalar field should not interfere with the thermal history of the Universe. It should be ``invisible" for much of its evolution after inflation and only start dominating around the present epoch and precipitate the late-time acceleration. Naturally, the construction of such a model is a very challenging endeavour, because it has to explain simultaneously the observations of both inflation and dark energy, but many successful models do exist \cite{Peloso:1999dm, Ng:2001hs, Dimopoulos:2000md, Dimopoulos:2001ix, Majumdar:2001mm, Nunes:2002wz, Dimopoulos:2002hm, Giovannini:2003jw, Spokoiny:1993kt, Joyce:1997fc, Sami:2004xk, Rosenfeld:2005mt, Pallis:2005hm, Pallis:2005bb, Brax:2005uf, Cardenas:2006py, BuenoSanchez:2006epu, BuenoSanchez:2006fhh, Rosenfeld:2006hs, Gomez:2008js, Bastero-Gil:2009wdy, Chiba:2012cb, Hossain:2014xha, Hossain:2014coa, Hossain:2014ova, Geng:2015fla, Haro:2015ljc, deHaro:2016ftq, Guendelman:2016kwj, Agarwal:2017wxo, Ahmad:2017itq, Geng:2017mic, Rubio:2017gty, Dimopoulos:2017zvq, Dimopoulos:2017tud, Akrami:2017cir, Bettoni:2018utf, Dimopoulos:2018eam, Haro:2018jtb, Durrive:2018quo, Bettoni:2018pbl, Haro:2018zdb, Wetterich:2019qzx, Dalianis:2019asr, Hashiba:2019mzm, Dimopoulos:2019ogl, Ahmad:2019jbm, Kleidis:2019ywv, Bettoni:2019dcw, Dimopoulos:2019gpz, Benisty:2020vvm, Cases:2020kvc, Benisty:2020qta, Arbey:2020ldf, Gangopadhyay:2020bxn, Dimopoulos:2020pas, Es-haghi:2020oab, AresteSalo:2021lmp, AresteSalo:2021wgb, Salo:2021vdv, Karciauskas:2021fdu, Bettoni:2021qfs} (for recent reviews see Refs.~\cite{Jaman:2022bho,deHaro:2021swo}).

First of all, one needs to construct an inflationary phase with a successful exit. 
The model must bridge the enormous gap of
energy density between inflation and dark energy, in a way that introduces as little fine-tuning as possible, for this is the main motivation for quintessence over $\Lambda$CDM. Indeed, a successful quintessential inflation model needs to rely on realistic theoretical foundations. Additionally, in quintessential inflation an
alternative reheating method is needed since the  scalar field  must survive until late times to become quintessence. Therefore, conventional reheating through inflaton decay is not applicable. Fortunately, a number of successful reheating mechanisms exist, such as instant preheating \cite{Felder:1998vq, Campos:2002yk}, curvaton reheating \cite{Feng:2002nb, BuenoSanchez:2007jxm} or Ricci reheating \cite{Dimopoulos:2018wfg,
Opferkuch:2019zbd, Bettoni:2021zhq}.  

In the post-inflationary era till the present epoch, the scalar potential should be steep, allowing  the radiation domination to commence, followed by the thermal history as envisaged by hot big bang. The steep potential is necessary for sending the field into hiding after the end of inflation. In particular, the post-inflationary dynamics is characterized by a field that evolves in the kinetic regime for some time~\cite{Spokoiny:1993kt, Joyce:1997fc}, but it then overshoots the background and gets frozen due to Hubble damping. As the background energy density redshifts and becomes comparable to the field energy density, the field resumes its evolution. Today, in most scenarios, the scalar field slow-rolls down its flat potential while dominating the Universe again and driving the current accelerated expansion.
However, if we consider interaction of the scalar field with matter at present, the dark energy period is more complicated.

In this paper, we study a model of quintessential inflation in the context of $R^2$ Palatini gravity where the scalar field has a running non-minimal coupling to gravity. Employing Palatini gravity to study quintessential inflation was first considered in Ref.~\cite{Dimopoulos:2020pas}, considering a variation of the original quintessential inflation model in Ref.~\cite{Peebles:1998qn}. This toy-model investigation demonstrated that modeling quintessential inflation with Palatini gravity is promising. In this, much more elaborated and realistic work, we consider a simple negative exponential potential in the Jordan frame. When we transform the theory to the Einstein frame, the potential becomes flat for both negative and positive field values with a steep transition region in-between, resembling a step function. The two flat regions are suitable for inflation and quintessence. Working in the Palatini formulation allows us to modify the inflationary plateau, in particular, through the $R^2$ term. The running non-minimal coupling allows us to obtain the correct quintessence behaviour. To study the full time evolution of the system throughout its cosmic history, we provide the equations of motion of the scalar field and an ideal fluid component representing other matter sources in the universe. We solve these equations numerically and scan over the parameter space, finding working scenarios matching both the CMB and late-time observations for parameter values that are free of fine-tuning. A preliminary study of the model can be seen in Ref.~\cite{Dimopoulos:2022tvn}; here, our treatment and findings are more complete and comprehensive.

The paper is structured as follows. In the next section, we describe our model and perform the Jordan to Einstein frame transformation. Then, in Sec.~\ref{sec:history}, we describe the model's time evolution in a cosmological setup. We employ the slow-roll approximation and discuss the inflationary behaviour of the model, adopt Ricci reheating as the mechanism responsible for reheating the Universe and describe its details, and outline the post-inflationary expansion history, namely, kination, radiation/matter domination, and quintessence. Numerical results for inflationary and late-universe observables are presented in Sec.~\ref{sec:numerical_results}, and we conclude in Sec.~\ref{sec:conclusions}. Further computational details are relegated to the appendices.

\section{Setup}
\label{sec:setup}
In this section, we first present the action of the model in the Jordan frame. After a frame transformation we bring the action to its Einstein frame form. Then, we compute the equations of motion in both Jordan and Einstein frames and show how one can easily transition between them.
\subsection{The model}
\label{sec:model}
We start by considering the action in the Palatini formalism
\begin{equation} \label{eq:S_Jordan}
    S = \int \dd^4 x \sqrt{-g} \qty[ \frac{\m^2}{2}F\qty(\varphi,R) - \frac{1}{2}g^{\mu\nu}\partial_\mu\varphi\partial_\nu\varphi - V(\varphi) ]+S_{\text{m}}[g_{\mu\nu},\psi] \, ,
\end{equation}
where $\m$ is the reduced Plank mass, $\psi$ collectively represents the matter fields other than the inflaton $\varphi$, and we take them to behave as an ideal fluid\footnote{Note that the matter action does not depend on the (independent) connection. This condition is necessary for the covariant conservation of the energy-momentum tensor.}.  The function $F(\varphi,R)$ takes the form
\begin{equation} \label{eq:F}
    F(\varphi,R)=\qty(1 + \frac{\xi}{\m^2} \varphi^2 )R+\frac{\alpha}{2\m^2}R^2 \, .
\end{equation}
We let the non-minimal coupling $\xi$ run as
\begin{equation} \label{eq:xi_running}
    \xi (\varphi) = \xi_{*} \left[ 1 + \beta \ln \left( \frac{\varphi^2}{\mu^2} \right) \right] \ ,
\end{equation}
with $\xi_{*} > 0$ and $\beta < 0$ constants, and $\mu$ an arbitrary reference scale.

In the Palatini formalism, the connection $\Gamma$ is independent of the metric $g_{\mu\nu}$. The connection features in the Ricci tensor, which is a function of the connection $\Gamma$ only, with
\begin{equation} \label{eq:R_in_Gamma}
    R=g^{\mu\nu}R_{\mu\nu}(\Gamma) \, .
\end{equation}
The form of the connection is determined by constraint equations obtained by varying the action with respect to $\Gamma$, and, in the presence of the non-minimal gravitational physics introduced by the non-zero $\xi$ and $\alpha$, it will differ from the standard Levi-Civita form.

The real scalar field $\varphi$, which plays the role of the inflaton and quintessence in quintessential inflation, is governed by an exponential potential
\begin{equation}
    V(\varphi)=M^4\text{e}^{-\kappa\varphi/\m} \, .
\end{equation}
The exponential form is well-motivated in particle physics (it usually appears in string theory and supergravity models, e.g. in gaugino condensation \cite{Gorlich:2004qm, Haack:2006cy, Lalak:2005hr}). It can produce quintessence in agreement with observations in its flat tail at $\varphi > 0$, and it is also suitable for quintessence from a theoretical point of view: we do not introduce a fine-tuned cosmological constant by hand, but instead $V \to 0$ for large $\varphi$, and the late time dark energy density arises dynamically from the equations of motion.

The action \eqref{eq:S_Jordan} is dynamically equivalent (as long as $\partial^2_{\chi}F\neq 0$) to 
\begin{equation}
    S=\int \dd^4 x \sqrt{-g}\qty[F(\varphi,\chi)+\partial_{\chi}F(\varphi,\chi)(R-\chi) - \frac{1}{2}g^{\mu\nu}\partial_\mu\varphi\partial_\nu\varphi - V(\varphi)]+S_{\text{m}}[g_{\mu\nu},\psi] \, ,
    \label{adjfbansdf}
\end{equation}
as can be seen by obtaining the equation of motion for the auxiliary field $\chi$ and plugging it back in Eq. \eqref{adjfbansdf}. Using this, the action can be cast in the form
\begin{equation}
    S=\int \dd^4 x \sqrt{-g}\qty[\frac{\m^2}{2}\qty(1 + \frac{\xi}{\m^2} \varphi^2 +\frac{\alpha}{\m^2}\chi)R-\frac{\alpha}{4}\chi^2- \frac{1}{2}g^{\mu\nu}\partial_\mu\varphi\partial_\nu\varphi - V(\varphi)]+S_{\text{m}}[g_{\mu\nu},\psi] \, .
\end{equation}
As is standard, we employ a conformal transformation (note that, in the Palatini formalism, this does not change $\Gamma$)
\begin{equation}
    g_{\mu\nu}\rightarrow \bar{g}_{\mu\nu}=\Omega^2 g_{\mu\nu}\equiv \qty(1 + \frac{\xi}{\m^2} \varphi^2 +\frac{\alpha}{\m^2}\chi)g_{\mu\nu}
\end{equation}
to express the action in the Einstein frame where the gravitational part takes the standard Einstein--Hilbert form:
\begin{equation} \label{eq:S_Einstein_varphi}
    S=\int \dd^4 x \sqrt{-\bar{g}}\qty[\frac{\m^2}{2}\bar{R}-\frac{1}{2}\frac{\m^2(\bar{\partial}\varphi)^2}{\qty(\m^2+\xi\varphi^2+\alpha\chi)}-\frac{\m^4\qty(V(\varphi)+\frac{\alpha}{4}\chi^2)}{\qty(\m^2+\xi\varphi^2+\alpha\chi)^2}]+S_{\text{m}}[\Omega^{-2}\bar{g}_{\mu\nu},\psi] \, .
\end{equation}
Note that, essentially, $\Omega^2 = \partial_R F(\varphi, R)$. We have introduced the short-hand notation $(\bar{\partial}\varphi)^2 \equiv \bar{g}_{\mu\nu}\bar{\partial}^\mu\varphi\bar{\partial}^\nu\varphi$, where $\bar{\partial}$ denotes a derivative with respect to the Einstein frame coordinates. Throughout the paper, we will use an overbar to denote Einstein frame quantities. Due to the standard form of the gravity sector, we will interpret all the usual cosmological observations in the Einstein frame.

To make the calculations that follow less cluttered, we define
\begin{equation} \label{eq:h}
    h(\varphi)\equiv \m^2+\xi\varphi^2 \, .
\end{equation}
We then get rid of the auxiliary field by obtaining its equation of motion. Let us, for a moment, ignore all matter except for the inflaton; then, we have
\begin{equation} \label{eq:chi_eq}
    \frac{\var{S}}{\var{\chi}}=0 \quad \Leftrightarrow \quad \chi=\frac{4\m^2V+h(\varphi)(\bar{\partial}\varphi)^2}{h(\varphi)\m^2-\alpha(\bar{\partial}\varphi)^2} \, ,
\end{equation}
giving
\begin{equation} \label{eq:Omega2}
    \Omega^2=\frac{h^2+4\alpha V}{h\m^2-\alpha (\bar{\partial} \varphi)^2} \, .
\end{equation}
Plugging both expressions back into the action gives \cite{Enckell:2018hmo, Antoniadis:2018ywb}
\begin{equation} \label{eq:actionnoncanonicaleinstein}
    S=\int \dd^4 x \sqrt{-\bar{g}}\qty[\frac{\m^2}{2}\bar{R}-\frac{1}{2}\frac{(\bar{\partial}\varphi)^2h\m^2}{h^2+4\alpha V} + \frac{\alpha}{4}\frac{(\bar{\partial}\varphi)^4}{h^2+4\alpha V}-\frac{V\m^4}{h^2+4\alpha V}]
    \, .
\end{equation}
Note that, because we were able to get rid of the non-dynamical auxiliary field through its equation of motion, the above action contains only one scalar field. This is in contrast to the metric version of the theory, where the auxiliary field is dynamical and the Einstein frame action contains two fields. 

The field can be made canonical via the redefinition
\begin{equation} \label{eq:field_transform}
    \frac{\dd \phi}{\dd \varphi}=\sqrt{\frac{h(\varphi)\m^2}{h(\varphi)^2+4\alpha V(\varphi)}} \, .
\end{equation}
Note that for large negative $\varphi$, this gives $\dd \phi/\dd \varphi \propto e^{\kappa\varphi/(2\m)}$, which, after integration, shows that $\phi$ approaches a constant as $\varphi \to -\infty$. We choose this constant to be equal to zero, so that $\phi$ is restricted to take positive values.

\begin{figure}[h]
    \centering
    \begin{subfigure}[b]{0.49\textwidth}
         \centering
         \includegraphics[width=\textwidth]{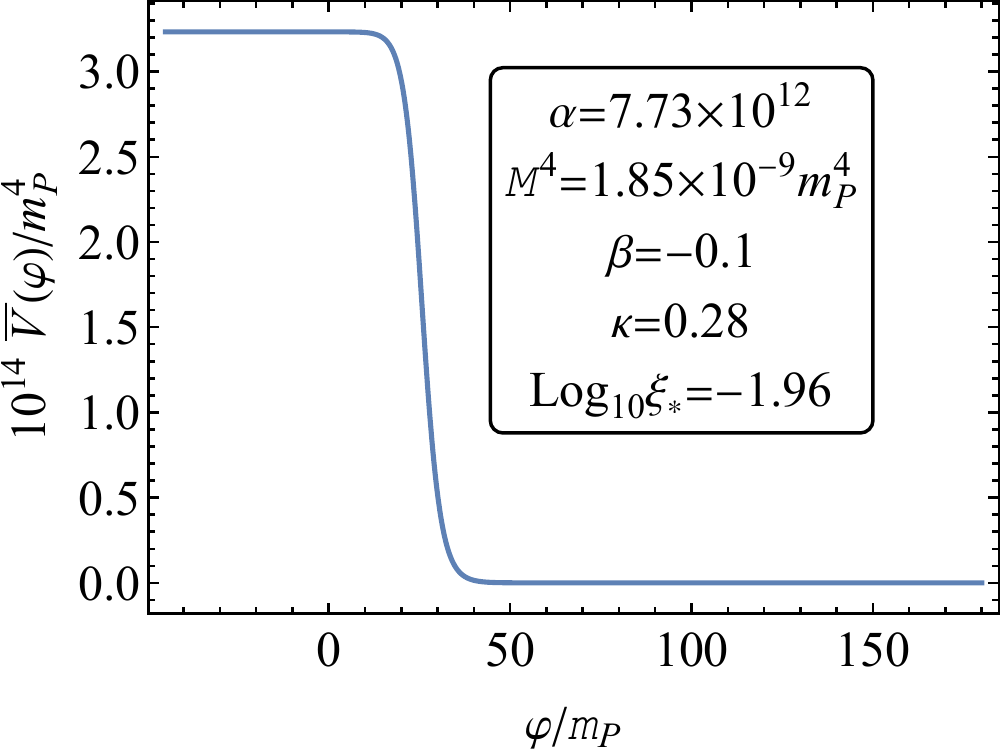}
     \end{subfigure}
     \begin{subfigure}[b]{0.49\textwidth}
         \centering
         \includegraphics[width=\textwidth]{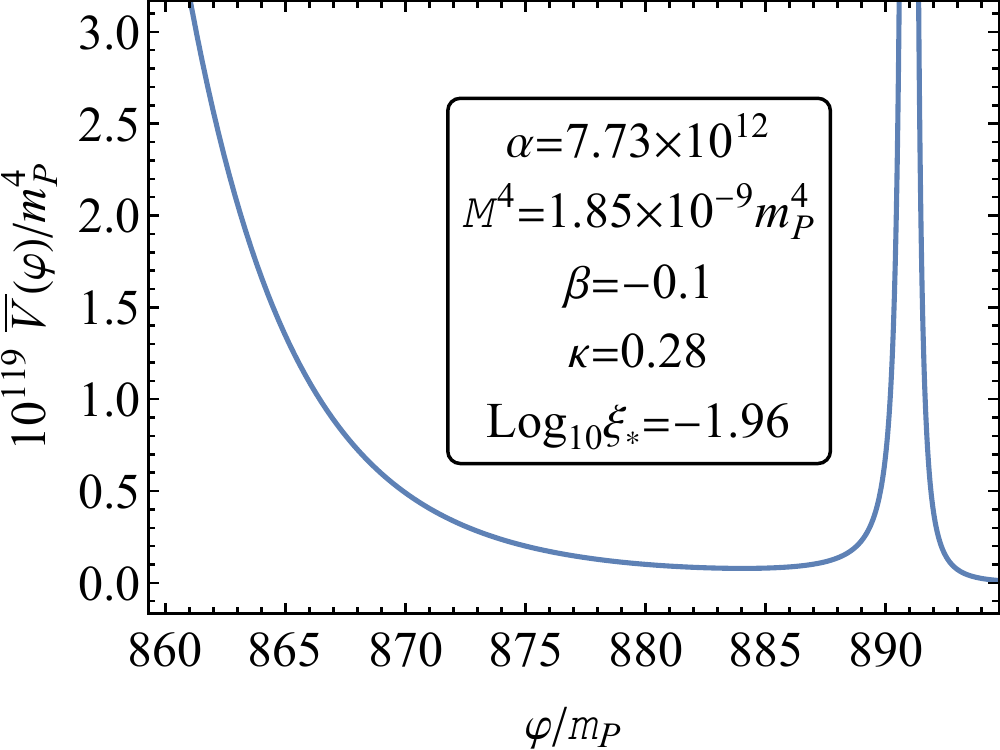}
     \end{subfigure}
    \caption{Potential in the Einstein frame $\bar{V}$ as a function of the field $\varphi$ (in Planck units), with the presented parameter values, in two regions: 
    around the inflation scale, $\varphi \sim 0$ (left), and around the point at which $1+\xi(\varphi)\varphi^2/\m^2$ becomes zero, \textit{i.e.}, $\varphi=890.99 \, \m$ (right). The height of the potential at this point is $\bar{V}(890.99)=1.14\times 10^{-94} \, \m^4$.}
    \label{Fig:examplepotential}
\end{figure}

The field redefinition leads finally to
\begin{equation} \label{eq:S_Einstein}
    S=\int \dd^4 x \sqrt{-\bar{g}}\qty[\frac{\m^2}{2}\bar{R}-\frac{1}{2}(\bar{\partial}\phi)^2 + \frac{\alpha}{4}\frac{h^2+4\alpha V}{h^2\m^4}(\bar{\partial}\phi)^4-\frac{V\m^4}{h^2+4\alpha V}]
    \, .
\end{equation}
Note the appearance of the higher-order kinetic terms for the scalar. As we will see below, they are negligible for most of cosmological evolution. Note also the form of the Einstein frame potential,
\begin{equation} \label{eq:V_Einstein}
    \bar{V}(\phi) \equiv \frac{V\m^4}{h^2+4\alpha V} = \frac{\m^4 M^4e^{-\kappa\varphi(\phi)/\m}}{(\m^2 + \xi\varphi(\phi)^2)^2 + 4\alpha M^4e^{-\kappa\varphi(\phi)/\m} } \, ,
\end{equation}
which chiefly determines the cosmological evolution of the model. An example case is depicted in Fig.~\ref{Fig:examplepotential}. The appealing features of the model are evident in the potential. For $\varphi< 0$, the potential decreases with increasing $\varphi$, but only slowly: the $\alpha$ term makes the potential flat and suitable for slow-roll inflation. For $\varphi>0$, the $\alpha$ term is subleading, and the potential decreases quasi-exponentially, modified by the change of variables \eqref{eq:field_transform}. The $\xi$ contribution modifies the potential; its running enables it to fix both the inflationary CMB observables and the late-time dark energy to values that match observations. For large enough $\varphi$, $\xi$ runs to negative values, causing $\bar{V}$ to first flatten and then start growing, forming a local minimum and a nearby peak when $1+\xi(\varphi)\varphi^2/\m^2$ becomes zero. For the parameters in Fig. \ref{Fig:examplepotential}, the zero occurs at $\varphi=890.99\m$, and at this point the height of the Einstein frame potential is $\bar{V}(890.99)=1.14\times 10^{-94}\m^4$ (notice the second term in the denominator in the potential regularizes the peak). Beyond the peak, the kinetic term in \eqref{eq:actionnoncanonicaleinstein} changes sign. In practice, as we will see below, dynamics never probe this region.

\subsection{Equations of motion in the Jordan frame}
\label{sec:Jordan_frame}

While the Einstein frame discussed in the previous section is useful for physical interpretation of the results, the equations of motion are easier to formulate in the Jordan frame, especially when we wish to include the non-inflaton matter contribution from \eqref{eq:S_Jordan} and thus go beyond the simplified Einstein frame action \eqref{eq:S_Einstein}. To obtain the equivalent of Einstein equations for our system (not a trivial task in the Palatini formulation with non-minimal gravity), we vary the action \eqref{eq:S_Jordan} with respect to the metric $g_{\mu\nu}$ and the connection $\Gamma$. The variation of the $F(\varphi,R)$ term reads
\begin{equation}
    \delta S=\frac{\m^2}{2}\int \dd^4 x \sqrt{-g} \qty[\qty(F_{R}(\varphi,R)R_{(\mu\nu)}-\frac{1}{2}g_{\mu\nu}F)\delta g^{\mu\nu}+(\partial_{R}F)g^{\mu\nu}\delta R_{\mu\nu}(\Gamma) ] \, ,
\end{equation}
where $F_{R}(\varphi,R)\equiv \partial_R F(\varphi,R)$ and parentheses around indices indicate the symmetric part of a tensor.
Meanwhile, the variation of the matter part gives the energy-momentum tensor, as usual:
\begin{equation}
    T_{\mu\nu}=-\frac{2}{\sqrt{-g}}\frac{\delta (S_{\varphi} + S_\text{m})}{\delta g^{\mu\nu}} = T^{(\varphi)}_{\mu\nu}+T^{(\text{m})}_{\mu\nu} \, .
\end{equation}
In the Jordan frame, the contributions from the field $\varphi$ and other matter components are independent from each other (indeed, this is the reason we work in the Jordan frame to begin with). We take the matter energy-momentum tensor to be of the ideal fluid form with energy density $\rho$ and pressure $p$, $T^{(\text{m})}_{\mu\nu}=({\rho}+p)u_{\mu}u_{\nu}+pg_{\mu\nu}$. We define the fluid's barotropic parameter as $w\equiv p/\rho$. The energy-momentum tensor of the field takes the standard form $T^{(\varphi)}_{\mu\nu} = \partial_\mu\varphi\partial_\nu\varphi - g_{\mu\nu}\qty(\frac{1}{2}(\partial\varphi)^2 + V)$. 

All in all, the variation of the action with respect to the metric gives
\begin{equation} \label{eq:Palatini_Einstein_eq}
    F_{R}R_{(\mu\nu)}-\frac{1}{2}g_{\mu\nu}F=\frac{1}{\m^2}T_{\mu\nu} \, .
\end{equation}
To make progress, we would like to express the left-hand side in terms of familiar geometric quantities. We start by taking the trace of \eqref{eq:Palatini_Einstein_eq},
\begin{equation} \label{eq:Palatini_Eisntein}
    F_{R}R-2F=\frac{1}{\m^2}T \, ,
\end{equation}
which algebraically relates the Palatini Ricci scalar to the matter sources. Using Eq. \eqref{eq:F}, this becomes
\begin{equation} \label{eq:R}
    R=-\frac{T}{\m^2+\xi\varphi^2} \, ,
\end{equation}
and the $R$-derivative of the $F$ function reads
\begin{equation} \label{eq:F_R}
    F_R=\left(1+\frac{\xi}{\m^2}\varphi^2\right)-\frac{\alpha T}{\m^4+\xi\m^2\varphi^2}\, .
\end{equation}
The trace of the energy-momentum tensor is
\begin{equation} \label{eq:energymomentumtrace}
    T=-g^{\mu\nu}\partial_\mu\varphi\partial_\nu\varphi-4V(\varphi)-\rho(1-3w) \, .
\end{equation}
Plugging these into \eqref{eq:Palatini_Einstein_eq}, we now have a relationship between $R_{(\mu\nu)}$ and the matter sources. However, as explained around \eqref{eq:R_in_Gamma}, $R_{\mu\nu}$ is defined through the Palatini connection $\Gamma$, which may differ from the Levi-Civita one, so we do not know how it relates to metric quantities such as the scale factor and the Hubble parameter. Varying the action with respect to the connection, we see (assuming symmetricity in the lower indices of the connection) that
\begin{equation}
    \delta R_{\mu\nu}=\tilde{\nabla}_{\lambda}\delta \Gamma^{\lambda}_{\nu\mu}-\tilde{\nabla}_{\nu}\delta \Gamma^{\lambda}_{\lambda\mu} \, ,
\end{equation}
where $\tilde{\nabla}$ is the covariant derivative defined with $\Gamma$. After integrating by parts in the action and some manipulations (relabelling indices and tracing over $\lambda$ and $\nu$), one obtains 
\begin{equation}
    \tilde{\nabla}_{\lambda}(\sqrt{-g}F_{R}g^{\mu\nu})=0 \, .
\end{equation}
This means that $\Gamma$ is actually the Levi-Civita connection of $\bar{g}_{\mu\nu}=(\partial_R F) g_{\mu\nu}$, that is,
\begin{equation}
    \Gamma^{\lambda}_{\mu\nu}=\frac{1}{2}\frac{1}{F_{R}}g^{\lambda \alpha}\left[\partial_{\mu}(F_{R}g_{\alpha\nu})+\partial_{\nu}(F_{R}g_{\alpha\mu})-\partial_{\alpha}(F_{R}g_{\mu\nu})\right] \, .
\end{equation}
Using this together with the Einstein equation \eqref{eq:Palatini_Einstein_eq}, we can derive an expression for the `metric' Einstein tensor defined by the Levi-Civita connection of the Jordan frame metric \cite{Sotiriou:2008rp}:
\begin{eqnarray} \label{eq:G}
    G_{\mu\nu}&&\equiv R_{\mu\nu}(g)-\frac{1}{2}g_{\mu\nu}R(g)
    =\frac{1}{\m^2F_{R}}T_{\mu\nu}-\frac{1}{2}g_{\mu\nu}\left(R-\frac{F}{F_{R}}\right)\nonumber
    \\
    &&+\frac{1}{F_{R}}(\nabla_{\mu}\nabla_{\nu}-g_{\mu\nu}\square)F_{R}
   -\frac{3}{2F_{R}^2}\left[\nabla_{\mu}F_{R}\nabla_{\nu}F_{R}-\frac{1}{2}g_{\mu\nu}(\nabla F_{R})^2\right] \, .
\end{eqnarray}
Here the covariant derivatives $\nabla$ are taken in terms of the Levi-Civita connection of the metric $g_{\mu\nu}$. In other words, we have eliminated the $\textit{a priori}$ independent connection $\Gamma$, which turns out to be just an auxiliary field. We see that the energy-momentum tensor is modified by rescalings and new effective matter sources.

Adopting now the flat FLRW metric, the metric Einstein tensor $G_{\mu\nu}$ can be written in terms of the scale factor $a$ and its time derivatives such as the Hubble parameter $H\equiv \dot{a}/ {a}$ in a standard way. Dot refers to a derivative with respect to the cosmic time.
The zeroth-zeroth component of Eq. \eqref{eq:G} then reads
\begin{equation} \label{eq:Palatini_Friedmann}
    3H^2=\frac{1}{\m^2 F_R}T_{00}+\frac{1}{2}\left(R-\frac{F}{F_R}\right)-\frac{3H\partial_0 F_R}{F_R}-\frac{3}{4F_R^2}\left(\partial_0 F_R\right)^2 \, ,
\end{equation}
while the $ij$ components read
\begin{equation} \label{eq:Palatini_Second_Friedmann}
    \dot{H}=-\frac{1}{2\m^2 F_R}(p+\rho+\dot{\varphi}^2)-\frac{\Ddot{F}_R}{2F_R}+\frac{3}{4F_R^2}(\dot{F}_R)^2 + \frac{H \dot{F}_R}{2F_R} \, ,
\end{equation}
where $R$ and $F_R$ are given in terms of the matter content by Eq. \eqref{eq:R} and \eqref{eq:F_R}
and
\begin{equation} \label{eq:T00}
    T_{00}=\rho+\frac{1}{2}\Dot{\varphi}^2+V(\varphi) \, .
\end{equation}
Eq.~\eqref{eq:Palatini_Friedmann} is a second-order algebraic equation for $H$, which can be solved in terms of the field and fluid variables $\varphi$, $\dot{\varphi}$, $\rho$, and $w$. A complication arises from the time derivatives of $F_R$: these contain also factors such as $\ddot{\varphi}$ and $\dot{\rho}$, which must be eliminated using the field and fluid equations introduced below. The procedure is explained in detail in Appendix \ref{sec:solving_H}.

For the fluid, it is shown in Ref. \citep{Koivisto:2005yk} that its energy-momentum tensor is conserved in this setup, so that
\begin{equation} \label{eq:matter_continuity_Jordan}
    \nabla_{\mu}T_{(\text{m})}^{\mu\nu}=0\quad \Rightarrow \quad \dot{\rho}+3H\rho(1+w)=0 \, .
\end{equation}
Finally, varying the action with respect to $\varphi$, we have
\begin{equation} \label{eq:phi_eom_Jordan_a}
    \Ddot{\varphi}+3H\dot{\varphi}+V'(\varphi)-\left(\xi(\varphi)+\frac{\xi'(\varphi)\varphi}{2}\right)\varphi R=0 \, ,
\end{equation}
which, using Eq. \eqref{eq:xi_running}, becomes
\begin{equation} \label{eq:phi_eom_Jordan}
    \Ddot{\varphi}+3H\dot{\varphi}+V'(\varphi)-\tilde{\xi}\varphi R=0 \, ,
\end{equation}
where we have defined
\begin{equation} \label{eq:xitilde}
    \tilde{\xi}\equiv\xi_{*}\left[1+\beta\left(1+\ln{\frac{\varphi^2}{\mu^2}}\right)\right].
\end{equation}
Equations \eqref{eq:matter_continuity_Jordan} and \eqref{eq:phi_eom_Jordan}, with \eqref{eq:R} and \eqref{eq:Palatini_Friedmann} for $R$ and $H$, form a complete set of equations from which the dynamics of the system can be solved.

\subsection{Between the Jordan and Einstein frames}
\label{sec:between_frames}

To give a physical interpretation for the dynamics, we want to relate the Jordan frame quantities to the Einstein frame ones. In both frames, we use a flat FLRW coordinate system with the metrics $g_{\mu\nu} = \mathrm{diag}(-1, a^2, a^2, a^2)$ and $\bar{g}_{\mu\nu} = \mathrm{diag}(-1, \bar{a}^2, \bar{a}^2, \bar{a}^2)$. We remind the reader that we use an overbar to denote the Einstein frame quantities. The spatial coordinates of these two flat FLRW coordinate systems match, but the time coordinates are rescaled.  As per our convention, we call the Jordan frame coordinates $x^\mu=(t,x^i)$ and the Einstein frame coordinates $\bar{x}=(\bar{t},x^i)$, and the conformal transformation gives for spacetime intervals
\begin{equation}
\begin{gathered}
    ds^2 = \dd x^\mu \dd x^\nu g_{\mu\nu} = \dd \bar{x}^\mu \dd \bar{x}^\nu \bar{g}_{\mu\nu}\Omega^{-2} \\
    \Leftrightarrow \, 
    (\dd x^i)^2 a^2 = (\dd x^i)^2 \bar{a}^2 \Omega^{-2} \, , \quad \, -\dd t^2 = -\dd \bar{t}^2 \Omega^{-2} \, , 
\end{gathered}
\end{equation}
where $\Omega^2 = F_R$ depends on time only. We obtain the relationships
\begin{equation} \label{eq:Einstein_vs_Jordan}
    \frac{\dd \bar{t}}{\dd t} = \sqrt{F_R} \, , \qquad \bar{a} = a \sqrt{F_R} \,
\end{equation}
as the master equations for moving between the two frames. With these, we can express various Einstein frame quantities in terms of the Jordan frame ones. In particular,
\begin{equation} \label{eq:HE}
    \frac{\dd}{\dd \bar{t}} \phi \equiv \dotE{\phi} = \frac{1}{\sqrt{F_R}}\frac{d \phi}{d \varphi}\dot{\varphi} \, , \qquad
     \bar{H} \equiv \frac{\dotE{\bar{a}}}{\bar{a}} = \frac{H}{\sqrt{F_R}} + \frac{1}{2}\frac{\dot{F_R}}{F_R^{3/2}} \, ,
\end{equation}
where a dot still denotes a derivative with respect to the Jordan frame time, and we introduced a circle over a symbol to indicate a derivative with respect to the Einstein frame time, so that $\dotE{x} = F_R^{-1/2}\dot{x}$.

The relation between the fluid energy-momentum tensors in the Jordan and Einstein frames is \citep{Dimopoulos:2020pas}
\begin{equation}
    \bar{T}_{\mu\nu}^{(\text{m})}=-\frac{2}{\sqrt{-\bar{g}}}\frac{\delta S_\text{m}}{\delta \bar{g}^{\mu\nu}}=-\frac{2}{\sqrt{-\bar{g}}}\frac{\partial g^{\alpha\beta}}{\partial \bar{g}^{\mu\nu}}\frac{\delta S_\text{m}}{\delta g^{\alpha\beta}}
    =\frac{\Omega^2}{\Omega^4}\left(-\frac{2}{\sqrt{-g}}\frac{\delta S_\text{m}}{\delta g^{\mu\nu}}\right)=\frac{1}{\Omega^2}T_{\mu\nu}^{(\text{m})} \, ,
    \label{adsiufubauisdfa}
\end{equation}
where we used $\partial g^{\alpha\beta}/\partial \bar{g}^{\mu\nu}=\Omega^2\delta^{\alpha}_{\mu}\delta^{\beta}_{\nu}$ and $\sqrt{-\bar{g}}=\Omega^4\sqrt{-g}$.

The Jordan frame ideal fluid is still ideal fluid in the Einstein frame; following Refs. \citep{Carroll:2003wy,Magnano:1993bd}, we write its energy-momentum tensor as
\begin{equation} \label{eq:energy_momentum_scaling}
    \bar{T}^{(\text{m})}_{\mu\nu}=(\bar{\rho}+\bar{p})\bar{u}_{\mu}\bar{u}_{\nu}+\bar{p}\bar{g}_{\mu\nu} \, , \qquad
    \bar{u}_{\mu}=\Omega u_{\mu} \, , \quad
    \bar{\rho}=\frac{\rho}{\Omega^4} \, , \quad
    \bar{p}=\frac{p}{\Omega^4} \, ,
\end{equation}
where the last equations relate the Jordan and Einstein frame quantities.
It follows that the barotropic parameter has the same expression in both frames:
\begin{equation}
    \bar{w}\equiv\frac{\bar{p}}{\bar{\rho}}=\frac{p}{\rho}=w \, .
\end{equation}
Below, we will always refer to the Einstein frame when talking of the barotropic parameter; we will omit the bar for simplicity of notation.

\subsection{Equations of motion in the Einstein frame}
\label{sec:Einstein_frame}
We are now ready to examine the Einstein frame equations of motion. Their full form is complicated---in the Einstein frame action \eqref{eq:S_Einstein_varphi}, the field and fluid components are coupled through the conformal factor $\Omega^{-2}$ inside $S_\text{m}$. In a general case with $\alpha \neq 0$, the fluid may even modify the $\chi$ constraint equation \eqref{eq:chi_eq} and, as a consequence, the field transformation \eqref{eq:field_transform}. We only present here approximate forms of the equations, free of some of these complications and valid during specific cosmological eras. Exact expressions can always be obtained by starting from the Jordan frame equations of Sec.~\ref{sec:Jordan_frame} and applying the transformations of Sec.~\ref{sec:between_frames}.

During inflation and right after it, the fluid is subdominant and can be ignored in the field equations. Varying the action \eqref{eq:S_Einstein} then gives \cite{Enckell:2018hmo}
\begin{equation} \label{eq:phi_eom_Einstein}
\begin{aligned}
    \qty[1+3\alpha\qty(1+\frac{4\alpha V}{h^2})\frac{\dotE{\phi}^2}{\m^4}]\ddotE{\phi} + 3\qty[1+\alpha\qty(1+\frac{4\alpha V}{h^2})\frac{\dotE{\phi}^2}{\m^4}] \bar{H} \dotE{\phi} & \\
    + 3\alpha^2\frac{\dotE{\phi}^4}{\m^4}\frac{\dd}{\dd\phi}\qty(\frac{V}{h^2}) + \frac{\dd}{\dd\phi}\bar{V} & = 0 \, ,
\end{aligned}
\end{equation}
with $h$ defined in \eqref{eq:h}.
The energy density and pressure of the field read \cite{Karam:2021sno}
\begin{equation} \label{eq:rho_p_Einstein}
\begin{aligned}
    \bar{\rho}_\phi &= \frac{1}{2}\qty[1+\frac{3}{2}\alpha\qty(1+\frac{4\alpha V}{h^2})\frac{\dotE{\phi}^2}{\m^4}]\dotE{\phi}^2 + \bar{V} \, , \\
    \bar{p}_\phi &= \frac{1}{2}\qty[1+\frac{1}{2}\alpha\qty(1+\frac{4\alpha V}{h^2})\frac{\dotE{\phi}^2}{\m^4}]\dotE{\phi}^2 - \bar{V} \, ,
\end{aligned}
\end{equation}
and the Hubble parameter can be written as $3\m^2 \bar{H}^2 = \bar{\rho}_\phi$. The higher-order kinetic terms are the only complication compared to a standard canonical scalar field.

At later times, the fluid becomes important, but the $\alpha$ terms turn out to be negligible. In this limit, the field transformation \eqref{eq:field_transform} can be solved explicitly to yield\footnote{Note that, when $\alpha=0$, we have $\phi \to -\infty$ as $\varphi \to -\infty$, contrary to the discussion below Eq.~\eqref{eq:field_transform}. When working in the $\alpha=0$ limit, we normalize the field so that $\phi=0$ when $\varphi=0$.}
\begin{equation} \label{eq:phi_tarnsform_solution}
    \sqrt{\xi}\varphi = \m \sinh(\sqrt{\xi}\phi/\m)\,,
\end{equation}
with the Einstein frame potential
\begin{equation} \label{eq:V_Einstein_approx}
    \bar{V}(\phi) = M^4 \frac{\text{exp}\left[-\frac{\kappa}{\sqrt{\xi}} \sinh(\frac{\sqrt{\xi}}{\m}\phi)\right]}{\cosh^4(\frac{\sqrt{\xi}}{\m} \phi)} \equiv \tilde{V}(\phi) \,.
\end{equation}
The field is coupled to the fluid; action \eqref{eq:S_Einstein} with the fluid contribution added in gives in the $\alpha \to 0$ limit:
\begin{equation} \label{eq:phi_eom_with_rho_mixing}
    \ddotE{\phi} + 3\bar{H} \dotE{\phi} + \frac{d \bar{V}}{d \phi} -\frac{1}{2F_R}\frac{d F_R}{d \phi}\qty(1-3w)\bar{\rho} = 0 \, ,
\end{equation}
where $F_R$ is given by \eqref{eq:F_R}, so that $\partial_\phi F_R/(2F_R) = \sqrt{\xi}\tanh\sqrt{\xi}\phi$, and we used
\begin{equation}
    \frac{1}{\sqrt{-\bar{g}}}\frac{\delta S_\text{m}}{\delta \phi}
    = \frac{\partial \Omega^2}{\partial \phi}\bar{g}^{\mu\nu}\qty(\frac{1}{\sqrt{-\bar{g}}}\frac{\delta S_\text{m}}{\delta g^{\mu\nu}})
    = -\frac{1}{\Omega^2}\frac{\partial\Omega^2}{\partial \phi}\bar{T}_m \, , \qquad \bar{T}_m = -(1-3w)\bar{\rho} \, . 
\end{equation}
Throughout the cosmic history, the fluid continuity equation in the Einstein frame can be obtained from the Jordan frame version \eqref{eq:matter_continuity_Jordan}, using the transformations of Sec.~\ref{sec:between_frames}. The result is
\begin{equation} \label{eq:matter_continuity_Einstein}
    \dotE{\bar{\rho}} + 3 \bar{H} \bar{\rho}(1+w) + \frac{1}{2F_R}\dotE{F}_R\qty(1-3w)\bar{\rho} = 0 \, .
\end{equation}
Multiplying the field equation \eqref{eq:phi_eom_with_rho_mixing} by $\dotE{\phi}$ gives the continuity equation for the field energy density. The inflaton-fluid coupling terms there and in \eqref{eq:matter_continuity_Einstein} are identical but have opposite signs: the coupling simply transfers energy from one component to the other. The coupling vanishes in the early universe when the fluid behaves like radiation, $w=1/3$, but it can be non-negligible during matter domination. We will discuss the effects of this coupling in more detail in Sec.~\ref{sec:numerical_results}.

Note that these expressions are still written partly in terms of the Jordan frame field $\varphi$, hidden in quantities like $h$, $V$, and $F_R$. In a general case, it is not possible to solve the field $\phi$ from $\varphi$ analytically. This is why, in our practical numerical computations, we work in the Jordan frame. The Einstein frame expressions of this section are for the benefit of developing a physical intuition of the system.

\section{Cosmic history with quintessential inflation} 
\label{sec:history}
Let us now turn to the time evolution of our model in a cosmological setup. In this section, we explore the cosmic history qualitatively through its many stages, starting from inflation and ending with quintessence domination. To make contact with the standard formalism discussed in the literature, we mostly work in the Einstein frame.

\subsection{Inflation}
\label{sec:inflation}
We start with the field at the plateau with $\varphi<0$ and high $\bar{V}$, with other matter components being negligible. We assume the high potential energy density dominates over the scalar's kinetic energy, giving rise to cosmic inflation, where the expansion of space accelerates. The plateau in $\bar{V}$ is suitable for \emph{slow-roll inflation}, where the field slowly moves towards positive values so that the potential gradient is balanced by Hubble friction. In this limit, the Einstein frame equations of motion \eqref{eq:phi_eom_Einstein} take the standard form
\begin{equation} \label{eq:sr_eom}
    3 \bar{H} \dotE{\phi} + \frac{\text{d}\bar{V}}{\text{d}\phi} = 0 \, , \qquad 3\bar{H}^2m_P^2 = \bar{V} \, ,
\end{equation}
where we neglected higher-order kinetic terms as subleading slow-roll corrections \cite{Enckell:2018hmo}. The evolution is characterized by the slow-roll parameters:
\begin{equation} \label{eq:sr_potential_variables}
    \epsilon_V \equiv \frac{1}{2}\qty(\frac{\text{d}\bar{V}}{\text{d}\phi}\frac{m_P}{\bar{V}})^2 \, , \qquad
    \eta_V \equiv 
    \frac{\text{d}^2\bar{V}}{\text{d}\phi^2}\frac{m_P^2}{\bar{V}} \, .
\end{equation}
For slow roll to be possible, we must have $\epsilon_V < 1$ and $|\eta_V| < 1$ at the corresponding field values. We can compute the slow-roll parameters for our potential \eqref{eq:V_Einstein} in the limit of constant $\xi$, that is, with $\beta=0$. To make the computation simpler, we use a result from \cite{Enckell:2018hmo} that relates $\epsilon_V$ and $\eta_V$ to their counterparts in the $\alpha=0$ limit (here $\tilde{\epsilon}$ and $\tilde{\eta}$, respectively). The results, by using Eq. \eqref{eq:field_transform} and the chain rule, can be expressed terms of the Jordan frame field $\varphi$ as
\begin{equation}
\begin{gathered} \label{eq:sr_V_in_model}
    \eta_V = \tilde{\eta} - 3\frac{4\alpha\tilde{V}}{1+4\alpha\tilde{V}} \, , \qquad
    \epsilon_V = \frac{\tilde{\epsilon}}{1+4\alpha\tilde{V}} \, ,
    \\ \tilde{\epsilon} = \frac{1}{2}\frac{\qty[\kappa\qty(1+\frac{\xi\varphi^2}{\m^2})+4\xi\frac{\varphi}{\m}]^2}{1+\frac{\xi\varphi^2}{\m^2}} \, , \quad
    \tilde{\eta}=\frac{7\kappa\xi\frac{\varphi}{\m}\left(1+\frac{\xi\varphi^2}{\m^2}\right)+\kappa^2\left(1+\frac{\xi\varphi^2}{\m^2}\right)-4\xi+16\xi^2\frac{\varphi^2}{\m^2}}{1+\frac{\xi\varphi^2}{\m^2}} \, ,
\end{gathered}
\end{equation}
and $\tilde{V}$ is defined in \eqref{eq:V_Einstein_approx}. The expression for $\epsilon_V$ reveals possible extrema with $\tilde{V}'=0$ at $\kappa\phi = -2 \pm \sqrt{4-\kappa^2/\xi}$. We demand that the potential is monotonic, i.e. $\tilde{V}'<0$ everywhere; this sets the restriction $\kappa^2 > 4\xi$ on the allowed parameter space.

Asymptotically, $\tilde{\epsilon}\sim \varphi^2$, diverging for both positive and negative $\varphi$. However, $\epsilon_V$ is suppressed by the exponential $\alpha\tilde{V}$ contribution so that $\epsilon_V \ll 1$ for $\varphi \ll -\m/\kappa$. This allows the system to undergo inflation even for large negative $\varphi$. Indeed, this was the motivation for us to introduce the $\alpha R^2$ term to our model \eqref{eq:S_Jordan} in the first place. The asymptotic behaviour $\eta_V \sim \tilde{\eta}\sim \varphi$ also reveals divergences for $|\varphi|\to\infty$, this time not removed by the $\alpha$ terms, making slow roll impossible for $\varphi \ll -m_P/(\kappa\xi)$. This leaves us with a range of field values near $\varphi=0$ that are compatible with slow-roll inflation. We start our inflationary evolution in slow-roll in this field range. As we will see in Sec.~\ref{sec:numerical_results}, typical values of the model parameters support the $60$ or so e-folds of inflation needed for a successful inflationary scenario.

The motivation for slow-roll inflation is that it produces a nearly scale-invariant spectrum of perturbations, compatible with the CMB observations \cite{Planck:2018jri, BICEP:2021xfz}
\begin{equation} \label{eq:CMB_observations}
    A_s = 2.1\times 10^{-9}\, , \quad n_s = 0.9649 \pm 0.0042\, , \quad \alpha_s = -0.0045 \pm 0.0067 \, , \quad r < 0.036 \, .
\end{equation}
Here $A_s$ is the scalar power spectrum amplitude, $n_s$ is the scalar spectral index, $\alpha_s$ its running, and $r$ is the tensor-to-scalar ratio at the CMB pivot scale $k_*=0.05\,\text{Mpc}^{-1}$. In the slow-roll limit, the perturbations can be computed using the standard formalism (see \textit{e.g.} \cite{Lyth:2009zz}), giving
\begin{equation} \label{eq:CMB_SR}
\begin{gathered}
    A_s = \frac{\bar{V}}{24\pi^2 m_P^4\epsilon_V} = \frac{\bar{H}^2}{8\pi^2 m_P^2 \epsilon_H} \, , \\
    n_s = 1 - 6\epsilon_V + 2\eta_V = 1 - 4\epsilon_H + 2\eta_H \, , \qquad r=16\epsilon_V=16\epsilon_H \, , 
\end{gathered}
\end{equation}
where we also gave the forms based on the Hubble slow-roll parameters,
\begin{equation} \label{eq:sr_hunnle_variables}
    \epsilon_H \equiv \frac{\dotE{\phi}^2}{2\bar{H}^2m_P^2} \approx \epsilon_V \, , \qquad\qquad \eta_H \equiv -\frac{\ddotE{\phi}}{\bar{H}\dotE{\phi}} \approx \eta_V - \epsilon_V \, ,
\end{equation}
where the approximations apply during slow roll. The expression for $\alpha_s$ depends on higher-order slow-roll parameters, which we omit for brevity; these can be found in \textit{e.g.} \cite{Lyth:2009zz}. Using the results \eqref{eq:sr_V_in_model}, we can also write down the full expression
\begin{equation} \label{eq:nsexplicit}
    n_s-1=-\kappa^2\qty(1+\frac{\xi\varphi^2}{\m^2})-10\xi\kappa\frac{\varphi}{\m}-8\xi\frac{1+2\frac{\xi\varphi^2}{\m^2}}{1+\frac{\xi\varphi^2}{\m^2}}\, .
\end{equation}

In our numerical results, we have $\beta \neq 0$, so the results \eqref{eq:sr_V_in_model}, \eqref{eq:nsexplicit} will be modified slightly. We will use these expressions as guidance when scanning over the parameter space, but we will compute the CMB observables from the Hubble slow-roll parameters as laid out in \eqref{eq:CMB_SR}. The modifications of inflation due to a non-zero $\beta$ turn out to be minor; $\beta$ is more important for the later evolution of the system, in particular, for fixing the final dark energy density.

\subsection{Kination}
\label{sec:kination}

Inflation ends when the field rolls down from the inflationary plateau to positive $\varphi$ values. As the field drops off the potential `cliff', see Fig.~\ref{Fig:examplepotential}, its velocity increases and the kinetic terms in the action \eqref{eq:S_Einstein} start to dominate over the potential. During this stage, the extra kinetic terms proportional to $\alpha\dotE{\phi}^4$ may play a role in the evolution. However, as the field velocity decreases due to Hubble friction, these terms die out quicker than the canonical $\dotE{\phi}^2$ kinetic term, which soon dominates. Analogously, in the Jordan frame, the $\alpha R^2$ term becomes subdominant compared to the linear $R$ term as the energy density of the universe, and thus its curvature,  decreases, and it stays subdominant until today. Thus, the $\alpha$ term is only important during and right after inflation.

After a transition period (lasting less than 10 e-folds according to the numerics of Sec.~\ref{sec:numerical_results}), the scalar field follows standard kination \cite{Spokoiny:1993kt,Joyce:1996cp,Joyce:1997fc,Pallis:2005hm,Pallis:2005bb,Gomez:2008js} with the equations of motion
\begin{equation} \label{eq:kination_eom}
    \ddotE{\phi} + 3 \bar{H} \dotE{\phi} = 0 \, , \qquad 3\bar{H}^2m_P^2 = \frac{1}{2}\dotE{\phi}^2 \, ,
\end{equation}
with the solution
\begin{equation} \label{eq:kination_solution}
    \dotE{\phi} \propto \bar{a}^{-3} \, , \qquad \bar{\rho}_\phi = \bar{p}_\phi = \frac{1}{2}\dotE{\phi}^2 \propto \bar{a}^{-6} \, .
\end{equation}
Note that the exponentially suppressed potential does not play a role during this stage. The evolution \eqref{eq:kination_solution} corresponds to a barotropic parameter $w=1$, which is quite distinct from the standard radiation or cold matter domination ($w=1/3$ and $0$, respectively). The period of kination leads to a non-standard expansion history of the universe, which, in particular, shifts the number of e-folds of inflation left at the Hubble exit of the CMB pivot scale $k=0.05 \text{Mpc}^{-1}$ from the standard 50--60 to 60--70.
We will return to this point in Sec.~\ref{sec:numerical_results}, where we match the CMB scale based on the full expansion history.

\subsection{Reheating} 
\label{sec:reheating}

In many conventional models of inflation, reheating occurs through the inflaton decaying into matter particles, which then take over the energy density and start the standard Hot Big Bang era. In quintessential inflation\footnote{and in general in non-oscillating inflation models}, the field condensate must be preserved and serve as dark energy later on. Therefore, radiation has to be created in some other way. There are many mechanisms which can facilitate this.

As an example, we consider one such mechanism, called Ricci reheating. Ricci
reheating was first considered by Ref.~\cite{Dimopoulos:2018wfg}. Then, it was refined first by
Ref.~\cite{Opferkuch:2019zbd}, which also coined the name, and further by Ref.~\cite{Bettoni:2021zhq}. In a nutshell,
the idea behind Ricci reheating is as follows. The
mechanism is based on the fact that, for a flat FRW Universe, the Ricci scalar (in the Einstein frame)
is \mbox{$\bar{R}=3(1-3w_\text{tot})\bar{H}^2$}, where $w_\text{tot}$ is the barotropic parameter of the whole Universe.
During slow-roll inflation we expect \mbox{$w_\text{tot}=-1$}, while after the end of inflation during kination we have \mbox{$w_\text{tot}=1$}. This implies that the sign of $\bar{R}$ changes
in the transition from inflation to kination. If one considers also a spectator scalar field $\psi$ with non-minimal coupling to gravity $\propto \bar{R}\psi^2$, then this change of sign in $\bar{R}$ would correspond to a change of sign in the effective mass-squared of $\psi$ generated due to the non-minimal coupling. Assuming that this effective mass-squared is positive during inflation, we can safely consider that the expectation value of $\psi$ is zero by the end of inflation. However, as we switch to kination, the effective mass of $\psi$ becomes tachyonic and the field is displaced from zero (which corresponds to a potential hilltop, after inflation) and begins oscillations in its effective potential. The oscillating $\psi$ has a particle interpretation and can decay into radiation, which eventually reheates the Universe, because its density is diluted less efficiently by the expansion than that of the free-falling inflaton during kination. 

The mechanism has a number of advantages compared to other reheating mechanisms considered in quintessential inflation. It can be very efficient, in contrast to gravitational reheating \cite{Ford:1986sy,Chun:2009yu}, which means it would not challenge Big Bang Nucleosynthesis (BBN); it does not require a coupling between the spectator field and the quintessential inflaton in an enhanced symmetry point, as would be the case of instant preheating \cite{Felder:1998vq,Campos:2002yk}; it does not need tuning of initial
conditions for the spectator field, as does the curvaton reheating mechanism \cite{Feng:2002nb,BuenoSanchez:2007jxm} and finally it does not presuppose a quintessential inflaton with dissipating properties as in warm quintessential inflation \cite{Dimopoulos:2019gpz} or the generation of primordial black holes \cite{Dalianis:2021dbs}. It only employs the fact that renormalisation in curved spacetime results generically in a non-minimal coupling of scalar fields to gravity. 

The additional Lagrangian density of the scalar field is
\begin{equation}
  \delta{\bar{\cal L}}=-\frac12\hat\xi \bar{R} \psi^2-
  \frac12 \bar{g}^{\mu\nu}\bar{\partial}_\mu\psi\bar{\partial}_\nu\psi-V(\psi)\,,
\label{dL}  
\end{equation}
where $V(\psi)$ is the part of the scalar potential which involves $\psi$ and
$\hat\xi$ is a non-perturbative coupling, which should not be confused with
$\xi$, the non-minimal coupling of the quintessential inflaton field.

Technically, the addition of the above in the Lagrangian density of the theory is yet another modification of gravity, which must be taken into account when switching between the Jordan and Einstein frames. However, we consider that \mbox{$\sqrt{\hat{\xi}}|\psi|\ll m_P$} always, which means that the influence of $\psi$ on gravity remains always negligible. Thus, in effect, we can consider that the only effect of the above non-minimal coupling is to provide a contribution to the effective mass-squared of the spectator field. Additionally, the
condition \mbox{$\sqrt{\hat{\xi}}|\psi|\ll m_P$} allows us to consider a perturbative scalar potential, which around the expectation value of the field during inflation, can be written as
\begin{equation}
V(\psi)=\frac12 m^2\psi^2+\frac14\lambda\psi^4+\cdots\,,
\label{Vpsi}
\end{equation}
where the ellipsis denotes higher-order non-renormalisable terms, presumed
negligible. We will consider at first that the non-minimal coupling overwhelms
the bare effective mass-squared \mbox{$|m^2|\ll|\hat\xi \bar{R}|$} so we can ignore
the first term on the right-hand-side above. This sets a limit on the mass which we discuss in Appendix~\ref{sec:spectator_mass}.
We will also consider a positive
perturbative self-coupling
\mbox{$0<\lambda<1$}, so that the potential is stabilised by the quartic term and not by
non-renormalisable terms, although a modification of our results in the latter
case is straightforward.

In Ref.~\cite{Bettoni:2021zhq} it was shown that after the end of inflation, the field $\psi$ oscillates as determined by the terms in \eqref{Vpsi}
that stabilize $V(\psi)$, 
while the effect of the central
potential hill (generated by the non-minimal coupling) is diminishing (and
negligible) because $\bar{R} \sim \bar{H}(t)$ is decreasing after inflation. If the stabilising
potential is quartic, as is the case of Eq.~(\ref{Vpsi}), then the density of
the oscillating condensate decays as radiation, \mbox{$\bar{\rho}_\psi\propto \bar{a}^{-4}$} \cite{Turner:1983he}. However, if the potential were stabilised by a non-renormalisable term,
this would not have been so. Fortunately, Ref.~\cite{Bettoni:2021zhq} demonstrated that it is
largely irrelevant which term stabilises the potential $V(\psi)$. This is
because in Ref.~\cite{Bettoni:2021zhq} it was shown that the primary reheating effect is not the perturbative
decay of the coherently oscillating $\psi$ condensate, but the non-perturbative
particle production on the hilltop, right after the end of inflation. At this
moment, the field finds itself on top of a potential hill, leading to ample production of radiation due to a tachyon instability. In Ref.~\cite{Bettoni:2021zhq},
it was claimed that the produced radiation dominates over the one corresponding
to the oscillating condensate. Because the latter is diluted (at least) as fast
as radiation, it never becomes important, at least as long as the quadratic
term in Eq.~(\ref{Vpsi}) remains negligible. These considerations simplify our treatment, because they suggest that radiation is
immediately produced at the end of inflation, and the further evolution of
the oscillating $\psi$ condensate is irrelevant. The only question is how much
radiation is produced. 

An estimate of the size of the spread of a scalar field condensate on top of a potential hill is given by \mbox{$\langle\psi^2\rangle\simeq |m_{\rm eff}|^2$} \cite{Copeland:2002ku}, where the effective mass squared in our case is \mbox{$m_{\rm eff}^2=-6\hat\xi \bar{H}^2$} during kination,
which takes place near the end of inflation. Therefore, the density of radiation at the end of inflation is
\begin{equation}
\bar{\rho}_r^{\rm end}=\frac12|m_{\rm eff}^2|\langle\psi^2\rangle\simeq 18\,\hat\xi^2 \bar{H}_{\rm end}^4 \, ,
\label{rhorend}
\end{equation}
where `end' denotes the end of inflation.
Thus, we obtain
\begin{equation}
 \Omega_r^{\rm end}= 
  \frac{\bar{\rho}_r^{\rm end}}{\bar{\rho}_\text{tot}^{\rm end}}\simeq
  \frac{18\,\hat\xi^2
  \bar{H}_{\rm end}^4}{3\bar{H}_{\rm end}^2m_P^2}=
  6\,\hat\xi^2\left(\frac{\bar{H}_{\rm end}}{m_P}\right)^2.
\label{rhoratio}
\end{equation}
During kination, the total density of the Universe decreases as
\mbox{$\bar{\rho}_\text{tot}\propto \bar{a}^{-6}$}, while for radiation we have
\mbox{$\bar{\rho}_r\propto \bar{a}^{-4}$}, which means that \mbox{$\bar{\rho}_r/\bar{\rho}_\text{tot}\propto \bar{a}^2$}.
Therefore,
\begin{eqnarray}
&&  \left.\frac{\bar{\rho}_r}{\bar{\rho}_\text{tot}}\right|_{\rm end}=
  \left(\frac{\bar{a}_{\rm end}}{\bar{a}_{\rm reh}}\right)^2
  \left.\frac{\bar{\rho}_r}{\bar{\rho}_\text{tot}}\right|_{\rm reh}\nonumber\\
  \Rightarrow &&
  \left(\frac{\bar{a}_{\rm end}}{\bar{a}_{\rm reh}}\right)^2\simeq
  6\,\hat\xi^2\left(\frac{\bar{H}_{\rm end}}{m_P}\right)^2,
\label{aratio}
\end{eqnarray}
where `reh' denotes reheating, which is the moment that radiation takes over
and we have \mbox{$\bar{\rho}_r\simeq\bar{\rho}_\text{tot}$}. The density of the Universe at reheating
is straightforward to find, by considering that \mbox{$\bar{\rho}_\text{tot}\propto \bar{a}^{-6}$}.
Indeed, we get
\begin{equation}
  \bar{\rho}_\text{tot}^{\rm reh}=\left(\frac{\bar{a}_{\rm end}}{\bar{a}_{\rm reh}}\right)^6\bar{\rho}_\text{tot}^{\rm end}
 \simeq 648\,\hat\xi^6\frac{\bar{H}_{\rm end}^8}{m_P^4}\,,
\label{rhoreh}
\end{equation}
where we used Eq.~(\ref{aratio}) and
\mbox{$\bar{\rho}_\text{tot}^{\rm end}=3\bar{H}_{\rm end}^2m_P^2$}. Therefore, using that at reheating
$\bar{\rho}_\text{tot}\simeq\bar{\rho}_r=\frac{\pi^2}{30}g_* T^4$, the reheating temperature is 
\begin{equation}
  T_{\rm reh}\simeq 6\left(\frac{15}{\pi^2g_*}\right)^{1/4}
  \hat\xi^{3/2}\,\frac{\bar{H}_{\rm end}^2}{m_P}\,,
\label{Treh}
\end{equation}
where $g_*$ is the number of effective relativistic degrees of freedom at
reheating.

The allowed reheating efficiency for successful reheating is
\begin{equation}
10^{-18}\lesssim\Omega_r^{\rm end}<1 \, . \label{Omegarange} 
\end{equation}
The lower bound in the 
range of the reheating efficiency in Eq.~(\ref{Omegarange})
is obtained from gravitational reheating 
\cite{Ford:1986sy,Chun:2009yu},
which challenges the process of BBN due to an over-enhancement of primordial gravitational waves during kination. Indeed, gravitational
reheating suggests \mbox{$\bar\rho_r^{\rm gr}\sim 10^{-2}\bar H_{\rm end}^4$}, where \mbox{$\bar H_{\rm end}\sim 10^{-8}m_P$} and we used that \mbox{$\bar V_{\rm end}^{1/4}\sim 10^{-4}m_P$}, as the numerical scans of Sec.~\ref{sec:numerical_results} give.
The value of $\bar{V}_{\rm end}^{1/4}$ roughly corresponds to $\m/(4\alpha)^{1/4}$, that is, $\bar{V}^{1/4}$ on the plateau during inflation. Using Eq.~(\ref{rhoratio}), we obtain the range of the non-minimal coupling of the spectator field 
\begin{equation}
    0.1\lesssim\hat\xi<10^8 \, .
    \label{hatxirange}
\end{equation}
This range includes values of \mbox{$\hat\xi\sim 1$}, which means that no fine-tuning is required for our mechanism to work.
A similar lower bound on $\hat\xi$ is obtained when considering the density of the primordial gravitational waves generated by inflation. Indeed, the density of the gravitational radiation at the end of inflation is 
\mbox{$\bar\rho_{_{\rm GW}}^{\rm end}\simeq\frac{1}{4\pi^2}\bar H_{\rm end}^4$}
(see appendix~\ref{rhoGW}).
We require that \mbox{$\bar\rho_r/\bar\rho_{_{\rm GW}}>20$} at BBN, but because both $\bar\rho_r$ and $\bar\rho_{_{\rm GW}}$ decrease with the expansion as $\bar a^{-4}$, we have the same requirement at the end of inflation. In view of Eq.~(\ref{rhorend}), this requirement becomes \mbox{$\hat\xi>1/6\sqrt 2\,\pi\simeq 0.038$}, which agrees with the range in Eq.~(\ref{hatxirange}).

Equations~(\ref{Treh}) 
and (\ref{hatxirange}) suggest that the reheating temperature ranges as
\begin{equation}
    10\;{\rm GeV}\lesssim T_{\rm reh} < 10^{14}\,{\rm GeV}\,.\label{Trehreange}
\end{equation}

\subsection{Radiation and matter domination}
\label{sec:radiation_matter_domination}

After reheating, the universe is dominated by hot radiation, and the barotropic parameter settles to $w=1/3$. As the univserse cools, particles in the thermal bath start to become non-relativistic, and this cold matter eventually takes over. We approximate this to happen instantaneously when $\bar{\rho}_r \simeq 10^{-110}m_P^4$, corresponding to a temperature of $\sim 0.8$ eV \cite{Lyth:2009zz}.

At the same time, the field follows the equation of motion \eqref{eq:phi_eom_with_rho_mixing}, veering away from kination once radiation starts to take over. While the fluid is relativistic, $w=1/3$, the field and fluid don't mix directly. However, in the presence of radiation the Hubble parameter is larger than it would be if induced by $\phi$ alone, and this increases the importance of the friction term. The field velocity $\dotE{\phi}$ starts to decrease dramatically, until the field essentially freezes to a near-constant value $\phi_\text{fr}$. Using the known scalings of the scalar and radiation energy densities, and assuming a negligible scalar potential, we can write
\begin{equation}
    \bar{\rho}_\text{tot} = \bar{\rho}_{\phi}^\text{kin}\qty(\frac{\bar{a}}{\bar{a}_{\text{kin}}})^{-6} + \bar{\rho}_{r}^\text{kin}\qty(\frac{\bar{a}}{\bar{a}_{\text{kin}}})^{-4} \, ,
\end{equation}
where `kin' refers to a moment at the beginning of standard kination with $\bar{\rho}_{r}^\text{kin}/\bar{\rho}_{\phi}^\text{kin}\equiv \Omega_{r}^\text{kin} \ll 1$. With this and $3\bar{H}^2m_P^2=\bar{\rho}_\text{tot}$, we can solve the frozen field value as
\begin{equation} \label{eq:delta_phi}
\begin{aligned}
    \phi_\text{fr}-\phi_\text{kin}
    &= \int_{\bar{t}_{\text{kin}}}^\infty \dd \bar{t} \, \dotE{\phi}
    = \int_{\bar{a}_{\text{kin}}}^\infty \dd \bar{a} \frac{\dotE{\phi}}{\bar{a} \bar{H}} \\
    &= \int_{\bar{a}_{\text{kin}}}^\infty \frac{\dd \bar{a}}{\bar{a}_{\text{kin}}} \frac{\sqrt{2\bar{\rho}_{\phi}^\text{kin}}(\bar{a}/\bar{a}_{\text{kin}})^{-4}}{\sqrt{\qty[\bar{\rho}_{\phi}^\text{kin}(\bar{a}/\bar{a}_{\text{kin}})^{-6} + \bar{\rho}_{r}^\text{kin}(\bar{a}/\bar{a}_{\text{kin}})^{-4}]/(3m_P^2)}} \\
    &= \sqrt{6}m_P\sinh^{-1}\qty(1/\sqrt{\Omega_{r}^\text{kin}})
    \approx \sqrt{6}\ln 2 - \sqrt{\frac{3}{2}}\ln \Omega_{r}^\text{kin} \, .
\end{aligned}
\end{equation}
As the kinetic energy of the field drops, the potential again starts to play an important role in field evolution, complicating the dynamics. Two basic behaviours emerge: the field may completely freeze, so that its potential energy comes to dominate over the kinetic one and the field's barotropic parameter becomes $-1$, or the field may start to follow a scaling attractor with slow time evolution \cite{Wetterich:1987fm,Ferreira:1997au,Copeland:1997et}. To estimate which fate is more likely, we can approximate the potential locally around $\phi=\phi_0$ with the exponential
\begin{equation} \label{eq:exp_pot_approx}
\begin{gathered}
    \bar{V}(\phi) \approx M^4_{\text{eff}}e^{-\kappa_\text{eff}\phi/m_P} \, , \\
    M^4_{\text{eff}} \equiv \frac{e^{-\frac{\kappa}{\sqrt{\xi(\phi_0)}}\sinh(\sqrt{\xi(\phi_0)}\phi_0/m_P) + \kappa\phi_0\cosh(\sqrt{\xi(\phi_0)}\phi_0/m_P)}}{\cosh^4(\sqrt{\xi(\phi_0)}\phi_0/m_P^2)} \, , \qquad
    \kappa_\text{eff} \equiv \kappa\cosh(\sqrt{\xi(\phi_0)}\phi_0/m_P) \, .
\end{gathered}
\end{equation}
If $\kappa_\text{eff}$ is approximately a constant, then $\kappa_\text{eff} < \sqrt{2}$ leads to freezing, and $\kappa_\text{eff} > \sqrt{2}$ gives the scaling solution. In our model in the examples below, we find $\kappa_\text{eff}$ to be small and slowly changing, leading indeed to a freezing behaviour.

After matter becomes non-relativistic with $w \neq 1/3$, time evolution is further complicated by the direct coupling between the fluid and the field in \eqref{eq:phi_eom_with_rho_mixing}, \eqref{eq:matter_continuity_Einstein}. In practice, the dynamics have to be solved numerically; we do this in Sec.~\ref{sec:numerical_results}.
 
\subsection{Quintessence domination}
\label{sec:quintessence_domination}

As the field rolls, $\xi$ from \eqref{eq:xi_running} runs to smaller and smaller values, and the Einstein frame potential \eqref{eq:V_Einstein} becomes flatter and flatter, becoming more suitable for quintessence with a slowly rolling field. Indeed, as mentioned in Sec.~\ref{sec:model}, eventually $\xi$ runs to negative values; around this point, the Einstein frame potential develops a local minimum and then starts to grow again, with a high positive peak near $\xi(\varphi)\varphi^2=-\m^2$. The coupling to matter can cause the field to overshoot the minimum and oscillate around it a few times, but eventually, as the fluid energy density dilutes away, the field will settle into the potential minimum at $\varphi\equiv\varphi_\text{fin}$. Its barotropic parameter $w_\phi=-1$ and its energy density, given by the height of the potential, become constant. The quintessence field then behaves as dark energy. To match observations, we need $\bar{V}(\varphi_\text{fin})= 7.23\times 10^{-121} m_P^4$, computed assuming that $\bar{H}=67.66 \text{km/s/Mpc}$ and that roughly 70\% of the energy density of the universe today is in dark energy. To be more precise, the dark energy fraction today is \cite{Planck:2018vyg}
\begin{equation} \label{eq:coincidence}
   \Omega_{\phi}=\Omega_{\mbox{\tiny DE}}=0.6889\pm 0.0056 \, .
\end{equation}
Since we live in the transition period where both dark energy and matter have non-negligible roles, the quintessence field is not necessarily completely frozen yet. In our numerical results, we demand that the barotropic parameter of the field today respects the observational bounds of the CPL parametrisation \cite{Chevallier:2000qy},
\begin{equation} \label{eq:CPL}
     w_\text{\tiny DE}=w_\text{\tiny DE}^{0}+w_\mathrm{a}\qty(1-\frac{\bar{a}}{\bar{a}_{0}}) \, , \qquad
     w_\mathrm{a} \equiv -\frac{\dd w_\text{\tiny DE}}{\dd \bar{a}}\Bigg|_{\bar{a}_{0}} \, ,
\end{equation}
where `0' refers to today, and the limits are \cite{Planck:2018vyg}
\begin{equation} \label{eq:w0obs}
  -1\leq w_\text{\tiny DE}^{0}<-0.95
  \qquad \text{and} \qquad
  w_\mathrm{a} \in [-0.55,0.03] \, .
\end{equation}

\section{Numerical results} 
\label{sec:numerical_results}

In this section, we explain the details concerning the numerical side of our work. As it is explained above, in order to numerically solve the dynamics of the system, we work in the Jordan frame. It is then straightforward to obtain the corresponding quantities in the Einstein frame, where our intuition applies, by following the discussion in Sec.~\ref{sec:between_frames}. To be more explicit, we need to solve for the scale factor $a(t)$, the inflaton field $\varphi(t)$ and the fluid density $\rho(t)$ (remember that, at a classical level, homogeneity and isotropy impose that the fields depend on time only), since every other quantity depends on these.
In principle this could be done by solving the system of ordinary differential equations given by Eqs. \eqref{eq:Palatini_Second_Friedmann}, \eqref{eq:matter_continuity_Jordan}, and \eqref{eq:phi_eom_Jordan}. However, the Hubble factor can be algebraically solved to be
\begin{equation}
    H=-\frac{A}{B+2F_R}+\frac{\sqrt{3F_R(4T_{00}+\alpha R^2)}}{3(B+2F_R)} \, ,
    \label{aoidsfbiasdf}
\end{equation}
where the specific forms of $A$ and $B$, as well as the details of the calculation, can be found in Appendix \ref{sec:solving_H}. For our current discussion it suffices to know that $A$ and $B$ depend on $\varphi(t)$ (and its first derivative) and $\rho(t)$ only. This means that the initial system of ordinary differential equations given by Eqs. \eqref{eq:phi_eom_Jordan}, \eqref{eq:Palatini_Second_Friedmann} and \eqref{eq:matter_continuity_Jordan} is reduced to Eqs. \eqref{eq:phi_eom_Jordan} and \eqref{eq:matter_continuity_Jordan}, where $H$ is given by Eq. \eqref{aoidsfbiasdf}. These are the equations that we numerically solve. 

It is also worth commenting on our choice of the Jordan frame over the Einstein frame. One obvious advantage of working in the Einstein frame is that the gravitational sector of the action is simply the Einstein--Hilbert term. However, were we to work in the Einstein frame, Eq. \eqref{eq:field_transform} would need to be solved and inverted in order to obtain $\varphi(\phi)$, to then be plugged back in the action \eqref{eq:actionnoncanonicaleinstein} in order to express all the quantities in terms of the canonical Einstein frame field $\phi$. Furthermore, the action in the Einstein frame features a quartic kinetic term and a coupling between the inflaton and the background matter fields through a conformal factor in the matter action. Although during inflation the matter action is zero, during the subsequent cosmological eras this extra coupling is present, complicating the setup. Likewise, the quartic kinetic term, which complicates the equations of motion even further, cannot be \textit{a priori} discarded (although after solving the dynamics it is found to be in general negligible, see Fig. \ref{Fig:contributionstoenergydensity}). All of these considerations outweigh the only hurdle in the Jordan frame: gravity is non-linear. As a matter of fact, due to working in the Palatini formalism, we can profit from further simplifications as the one explained above, where the Hubble factor can be algebraically solved in terms of the inflaton and the background fields. In this way, we find the solution of the system to be much more approachable in the Jordan frame than in the Einstein frame. Finally, it is important to keep in mind that, as we have mentioned, once the dynamics is solved in the Jordan frame it is straightforward to obtain the analogous quantities in the Einstein frame by following the discussion in Sec. \ref{sec:between_frames}.

\subsection{Initial conditions}

During the inflationary era the only existing field is the inflaton (even if some matter fields existed they would be inflated away), so that $\rho(t)=0$. Therefore, the only equation to solve for is Eq. \eqref{eq:phi_eom_Jordan} (with $H$ given by Eq. \eqref{aoidsfbiasdf}), which is a second order ordinary differential equation. Thus, only two initial conditions are needed, $\varphi(t_i)$ and $\dot{\varphi}(t_i)$. We choose $\varphi(t_i)$ sufficiently negative to capture all the possible evolution histories when scanning over the parameter space, while respecting the bound that imposes that the field should not be much smaller than $-\m/(\kappa\xi)$ (\textit{c.f.} Sec. \ref{sec:inflation}), for which slow-roll is not possible. This usually amounts to having $\varphi(t_i)\sim-30\,\m$ and as it can be seen from Fig. \ref{fig:parameterspacecomparison} (see also the discussion in Sec. \ref{sec:numericalresultsforinflation}), using a smaller value would be of no help, since the region of the parameter space compatible with observations restricts $\varphi(t_i)>-30\,\m$. Furthermore, for simplicity, since the field will eventually reach the slow-roll attractor, we choose $\dot{\varphi}(t_i)$ such that slow-roll is satisfied. Effectively this means neglecting the second order derivative in Eq. \eqref{eq:phi_eom_Jordan}. With $\varphi(t_i)$ fixed, this equation only depends on $\dot{\varphi}(t_i)$, for which we can (numerically) solve to obtain the initial value. 

The end of inflation gives way to kination. During this era, some reheating mechanism transfers the energy density of the inflaton to the particles of the SM, which are modelled in our setup by a perfect fluid with energy density $\rho(t)$. In this way, the last needed initial condition is the initial energy density of radiation, at the end of inflation, $\rho(t_\text{end})$. It can be found by the following simple calculation
\begin{equation}
    \text{e}^{\bar{N}}\equiv\frac{\bar{a}_{\text{end}}}{\bar{a}_{*}}=\frac{\bar{a}_{\text{end}}}{\bar{a}_0}\frac{\bar{a}_0}{\bar{a}_{*}}=\frac{\bar{a}_{\text{end}}}{\bar{a}_0}\frac{\bar{H}_{*}}{\bar{a}_{*}\bar{H}_{*}}=\frac{T_0}{T_{\text{end}}}k_{*}^{-1}\bar{H}_{*} \, ,
\end{equation}
where $*$ corresponds to the time at which the CMB pivot scale exits the horizon during inflation (with $k_* = \bar{a}_* \bar{H}_* = 0.05\ \text{Mpc}^{-1}$), ``end'' corresponds to the end of inflation, and `0' corresponds to the present time. We have also set $\bar{a}_{0}=1$ and made the approximation $\bar{a}\propto T^{-1}$ from the end of inflation until today, where $T$ is the temperature of radiation. Using Eq. \eqref{eq:Einstein_vs_Jordan}, we can relate the number of e-folds in the Einstein and Jordan frames as~\cite{Karam:2017zno, Racioppi:2021jai}
\begin{equation}
    \bar{N}=
    N+\frac{1}{2}\ln{\frac{F_R^{\text{end}}}{F_R^{*}}} \, .
\end{equation}
Thus,
\begin{equation}
    \text{e}^{N}=\frac{T_0}{T_{\text{end}}}k_{*}^{-1}\sqrt{\frac{F_R^{*}}{F_R^{\text{end}}}}\bar{H}_{*} \, ,
\end{equation}
where $T_0 \approx 2.7$K. The initial energy density of radiation at the end of inflation can be written as
\begin{equation}
    \bar{\rho}=\frac{\pi^2}{30}g_{*}T^4 \, ,
\end{equation}
where $g_*=106.75$ is the number of relativistic degrees of freedom. Relating $\bar{\rho}$ to $\rho$ via \eqref{eq:energy_momentum_scaling} and gathering the above results together, we get
\begin{equation} \label{eq:initialradiationdensity}
    \rho(t_\text{end})=(F_R^{*})^2\frac{\pi^2 g_{*}}{30}\left[\frac{T_0}{\text{e}^{N}}k_{*}^{-1}\bar{H}_{*}\right]^4 \, ,
\end{equation}
written in terms of quantities that are either known or fixed by inflation.
Note the cancellation of $(F^{\text{end}}_R)^2$ due to the extra factor of $F_R^2$ coming from expressing the energy density in the Jordan frame. It is important to mention that when scanning over the parameter space we require that $\bar{\rho}(t_\text{end})$ satisfies two bounds, the upper one such that the inclusion of the radiation fluid at the end of inflation is a small perturbation to the overall dynamics, \textit{i.e.}, $\Omega_r^{\text{end}}<0.1$, and the lower one corresponding to the gravitational reheating limit, which is the least efficient reheating mechanism. Thus, we impose $\bar{\rho}(t_\text{end})>\bar{\rho}_{\text{grav}}=qg_{*}(\bar{H}_{\text{end}})^4/(480\pi^2)\simeq 2.25\times 10^{-2}(\bar{H}_{\text{end}})^4$ \cite{kostasbook}, where we have introduced $q\sim1$ because the spectrum is not exactly thermal.

\subsection{The parameter space} \label{subsec:parameterspace}
The model has six parameters, namely $\kappa$, $\xi_{*}$, $\beta$, $\mu$, $\alpha$, and $M^4$. It would be computationally costly to perform a scan over such a six-dimensional space. However, there are some simplifications that allow us to reduce the dimensionality of the parameter space. 

The first thing to notice is the scaling law the model obeys. Indeed, let us rescale the coordinates, background density and parameters in the Jordan frame as\footnote{Note that here $\lambda$ is just a constant factor and should not be confused with the $\psi$ field self-coupling in Eq. \eqref{Vpsi}.}
\begin{equation}
    x^{\mu}\rightarrow \lambda x^{\mu},\quad \rho\rightarrow \lambda^{-2}\rho,\quad \alpha\rightarrow \lambda^2 \alpha\quad \text{and}\quad M^4\rightarrow \lambda^{-2}M^4 \, . \label{eq:rescaling}
\end{equation}
From Eq. \eqref{eq:R} it immediately follows that under this transformation the Ricci scalar scales as 
\begin{equation}
    R\rightarrow \lambda^{-2}R \, .
\end{equation}
Likewise, from Eq. \eqref{eq:S_Jordan}, the action scales as 
\begin{equation}
    S\rightarrow \lambda^2 S \, .
    \label{eq:actionrescaling}
\end{equation}
Of course, the equations of motion are invariant under such a rescaling of the action. Furthermore, the quantity $\alpha M^4$ is also invariant. Looking at the expressions for the inflationary observables in Eq. \eqref{eq:sr_hunnle_variables}, one can see that the parameters $\alpha$ and $M^4$ only enter the expressions for $n_s$ and $r$ through the combination $\alpha M^4$, \textit{i.e.}, they are invariant under the rescaling \eqref{eq:rescaling}. It is not so for $A_s$, where $M^4$ enters its expression alone. 

From this discussion we conclude that it is enough to scan over the quantity $\alpha M^4$. For each value of $\alpha M^4$ we can fix $M^4$ such that $A_s$ satisfies the observational requirements from \eqref{eq:CMB_observations}. In this way we have reduced the dimensionality of the parameter space to five.

There is one extra simplification that can be made by taking into account that $\mu$ in Eq. \eqref{eq:xi_running} is an arbitrary scale that can be changed by reabsorbing it into $\xi_{*}$. Therefore it can be chosen to take the most convenient value, which, for us, is the field value at which the cosmological scales leave the horizon, $\varphi_*$. This way, around this scale the effect of the running is minimal and the non-minimal coupling is roughly just $\xi_{*}$. The dimensionality of the parameter space is now four. 

Having defined the degrees of freedom of the system, \textit{i.e.}, $\varphi(t)$, $a(t)$ and $\rho(t)$, the initial conditions, \textit{i.e.}, $\varphi(t_i)$, $\dot{\varphi}(t_i)$, and $\rho(t_\text{end})$, and the parameters over which to scan, \textit{i.e.}, $\kappa$, $\alpha M^4$, $\xi_{*}$ and $\beta$, we first focus on the inflationary regime of the theory. In this way, we start with the initial conditions discussed above and numerically solve the system until the end of inflation, defined by the condition\footnote{Note that the first slow-roll parameter in Eq. \eqref{eq:sr_V_in_model} is only an approximation. In the numerical study we also take into account the presence of the running in $\xi$.} $\epsilon_H\equiv \dotE{\phi}^2/(2\bar{H}^2\m^2)=1$. We take discrete slices in $\alpha M^4$, ranging from $0.0143$ to $1.43\times 10^6$ in steps of factor 10 and a region in $\beta$ around the central value of $-0.1$ with a resolution of $10^{-3}$ and scan over the parameters $\kappa$ and $\log_{10}\xi_{*}$ with values in the intervals $[0.2,0.7]$ and $[-2.5,-0.9]$, respectively, with resolutions of $5\cross 10^{-3}$. The reason behind choosing such a central value for $\beta$ is that we have found that a correct behaviour for quintessence is strongly peaked around it. 

As the values for the field and its velocity at the end of inflation will serve as the initial conditions for the beginning of the next cosmological era, we impose a set of conditions on the points obtained from the scan through which we obtain the valid region in the parameter space. In addition to fixing $A_s=2.1\times10^{-9}$ as discussed above, we require that:
\begin{itemize}
    \item The value of the scalar spectral index is equal to the central value obtained by Planck \citep{Aghanim:2018eyx}, \textit{i.e.}, $n_s=0.9649$.
    \item The value of the tensor-to-scalar ratio is within the latest observational bounds \citep{BICEP:2021xfz}, \textit{i.e.}, $r<0.036$.
    \item The value of the running of the scalar spectral index is within the $2\sigma$ bounds obtained by Planck \citep{Aghanim:2018eyx}, \textit{i.e.}, $-0.0179<\alpha_s<0.0089$.
    \item The initial energy density of radiation at the end of inflation, obtained via Eq. \eqref{eq:initialradiationdensity}, amounts to a small perturbation of the system, \textit{i.e.}, $\Omega_r^{\text{end}}<0.1$.
    \item The initial energy density of radiation at the end of inflation is larger than the energy density corresponding to gravitational reheating, \textit{i.e.}, $\bar{\rho}(t_\text{end})>2.25\times10^{-2}(\bar{H}^{\text{end}})^4$. 
\end{itemize}
The last two conditions translate to the available range in the number of e-folds from the time at which the cosmological scales exit the horizon until the end of inflation (see the right panel in Fig. \ref{fig:parameterspace1}). It is usually between $60$ and $75$. Also note that we have not imposed a correct value for the amplitude of the power spectrum $A_s$ as a condition since every single point in the parameter space already satisfies this, by exploiting the scaling property of the model explained above.

When inflation ends, and after imposing the above set of conditions to obtain the valid region of the parameter space, we use the final values of the field and its velocity as the initial conditions for the next cosmological era, as well as Eq. \eqref{eq:initialradiationdensity} for the radiation energy density, in order to solve Eqs. \eqref{eq:matter_continuity_Jordan} and \eqref{eq:phi_eom_Jordan}, with $H$ given by Eq. \eqref{aoidsfbiasdf}. The barotropic parameter of the fluid is of course $1/3$ up until the transition to the matter domination era, when it becomes $w=0$. We model this transition by a jump from $1/3$ to $0$ in the barotropic parameter of the background at the time when the energy density of radiation is equal to its value at matter-radiation equality, $\bar{\rho}_{\text{eq}}=1.27\times10^{-110} \m^4$ \cite{Aghanim:2018eyx}. The simulation is finished when the energy density ratio of the field, corresponding now to dark energy, is equal to the central value obtained by Planck \cite{Aghanim:2018eyx} of its value today, \textit{i.e.}, $\Omega_{\phi}^0=0.6889$. At this point we impose another set of conditions, which we list here.
\begin{itemize}
    \item The temperature of the universe at the onset of radiation domination is above $T_{\text{BBN}}\simeq 0.1 \text{MeV}$.
    \item The barotropic parameter of the field is within the latest observational bounds \citep{Aghanim:2018eyx}, \textit{i.e.}, $w_{\phi}^0<-0.95$.
    \item The running of the barotropic parameter of the field in the CPL parametrization is within the latest observational bounds \citep{Aghanim:2018eyx}, \textit{i.e.}, $-0.55<w_a^0<0.03$.
    \item The energy density of the field at present is within one order of magnitude from the central value obtained by Planck, $\bar{\rho}_{\text{DE}}^{\text{Planck}}=7.26\cross10^{-121}\m^4$ \cite{Aghanim:2018eyx}.
\end{itemize}

Importantly, we also consider the bound on the density parameter of gravitational waves coming from BBN constraints, $20\,\Omega_{_{\text{GW}}}^{\text{end}}<\Omega_r^{\text{end}}$, as discussed in Sec. \ref{sec:reheating}. By using Eq. \eqref{rhoratio} this bound translates to an allowed range of values for the non-minimal coupling between the reheaton and gravity $\hat{\xi}$.
The successful values of $\hat{\xi}$ for each point in the parameter space are presented in Table \ref{table}. The two points, for which
$\hat\xi$ is the largest ($\hat{\xi}\sim\mathcal{O}(1)$),
are highlighted in black in Fig. \ref{fig:deparameterspace} (while the rest are in red).

The points that satisfy this extra set of conditions are the successful points of our model. For them we have successful inflation, with correct inflationary predictions, as well as a correct evolution during the expansion history of the universe, with successful dark energy at the present time. 

\subsection{Numerical results for inflation} \label{sec:numericalresultsforinflation}
\begin{figure}[h]
     \centering
     \begin{subfigure}[b]{0.45\textwidth}
         \centering
         \includegraphics[width=\textwidth]{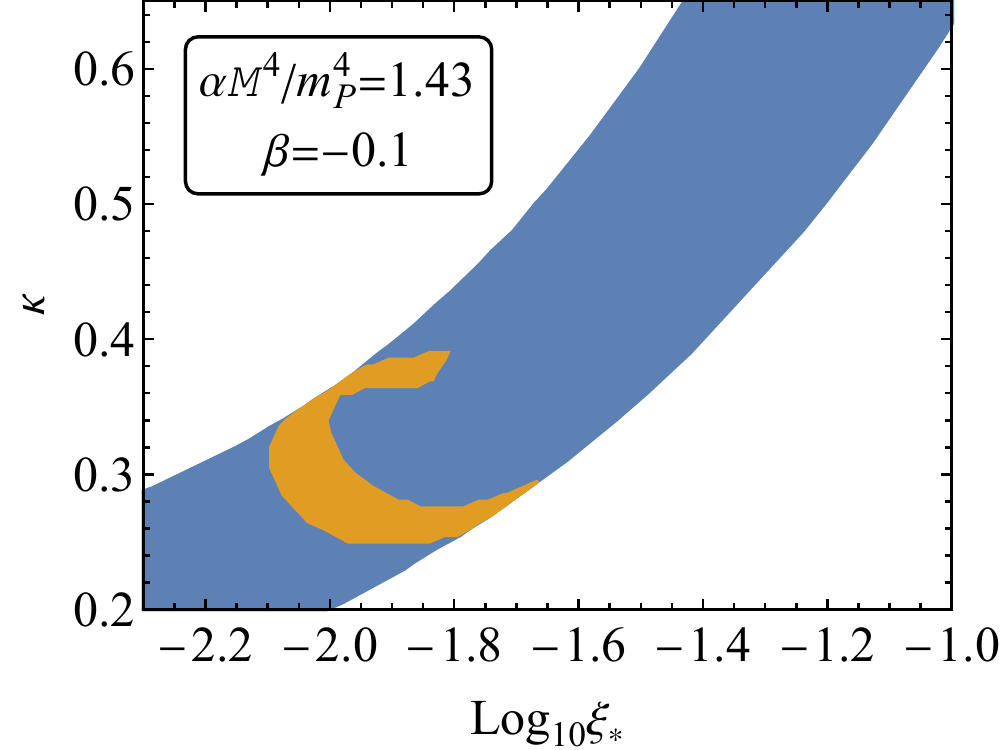}
     \end{subfigure}
     \begin{subfigure}[b]{0.45\textwidth}
         \centering
         \includegraphics[width=\textwidth]{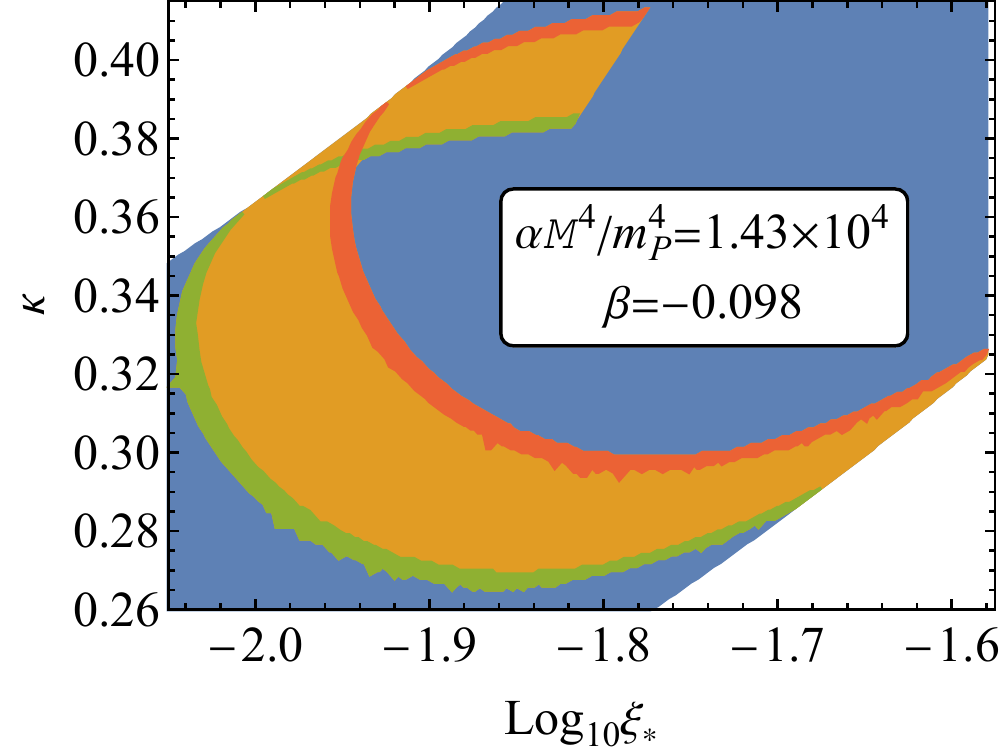}
     \end{subfigure}
     \caption{Left: Slice of the parameter space in the $(\log_{10}\xi_{*},\kappa)$ plane with $\beta=-0.1$ and $\alpha M^4/\m^4=1.43$. The blue points have a correct value of the scalar spectral index, while the orange points satisfy all observational constraints for inflation. Right: A zoomed-in slice with $\beta=-0.098$ and $\alpha M^4/\m^4=1.43\times 10^4$, depicting the bounds in parameter space corresponding to the bounds in the number of inflationary e-folds. The red region is close to saturating the gravitational reheating bound $\bar{\rho}(t_\text{end})>2.25\times10^{-2}(\bar{H}^{\text{end}})^4$ (which corresponds to the upper limit in the number of e-folds), while the green region is close to saturating the bound $\Omega_r^{\text{end}}<0.1$ (which corresponds to the lower limit in the number of e-folds).}
     \label{fig:parameterspace1}
\end{figure}

\begin{figure}[h]
    \centering
     \includegraphics[width=0.6\textwidth]{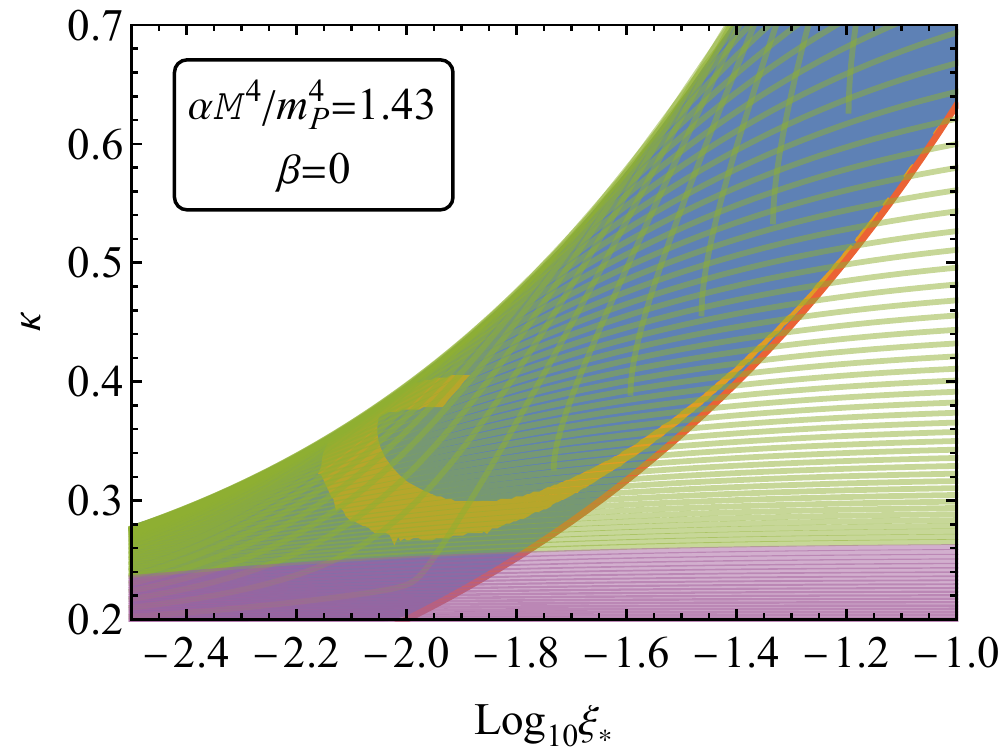}
     \caption{Slice of the parameter space in the $(\log_{10}\xi_{*},\kappa)$ plane with $\beta=0$ and $\alpha M^4/\m^4=1.43$, where we plot many curves $\kappa(\xi_{*})|_{n_s=0.9649}$ with $\varphi_{*}$ ranging from $-30\,\m$ to 0 (green) and from $-200\,\m$ to $-30\,\m$ (purple), as well as the curve $\kappa^2=4\xi(=4\xi_{*})$ (red), so that the condition for a monotonic potential $\kappa^2>4\xi(=4\xi_{*})$ is satisfied above it. The upper boundary of the parameter space coincides with the asymptotic upper bound from the green curves. Increasing $\varphi_{*}$ to more negative values explores a region of the parameter space that is not in agreement with observations, towards smaller and smaller $\kappa$, as can be seen from the purple curves. The parameter space of the theory lies between the asymptotic upper bound from the $\kappa(\xi_{*})|_{n_s=0.9649}$ curves and the condition $\kappa^2=4\xi(=4\xi_{*})$, as it should. }
      \label{fig:parameterspacecomparison}
\end{figure}

In this section, we present and analyze the obtained results for inflation. We remind the reader that the power spectrum strength at the pivot scale, $A_s$, is fixed to its observed value in all our results. In the left panel in Fig. \ref{fig:parameterspace1} we show an example slice of the parameter space in the $(\log_{10}\xi_{*},\kappa)$ plane with fixed $\beta=-0.1$ and $\alpha M^4/\m^4=1.43$. The blue points have a correct value for $n_s$ while the orange points satisfy the full set of conditions for inflation stated in Sec. \ref{subsec:parameterspace}. 
In order to understand the shape of the parameter space let us consider the $\beta=0$ case, for simplicity. First, we remember we have imposed the potential to be monotonic, \textit{i.e.}, $\kappa^2>4\xi=4\xi_{*}$. The lower boundary of the parameter space region corresponds to this requirement. The other consideration to take into account is Eq. \eqref{eq:nsexplicit}, which, since in the $\beta=0$ case the expressions for the slow-roll parameters in Eq. \eqref{eq:sr_V_in_model} are exact, is an exact expression for the scalar spectral index (in the slow-roll approximation). Since this is a quadratic equation in $\kappa$, it can be algebraically solved for, giving an expression depending on $\xi_{*}$ and $\varphi_{*}$ (the field at horizon exit), $\kappa=\kappa(\xi_{*},\varphi_{*})$. In Fig. \ref{fig:parameterspacecomparison} we plot in green the curve $\kappa(\xi_{*})|_{n_s=0.9649}$, for many values of the field at horizon exit, ranging from $-30\,\m$ to $0$ (in steps of $0.5\,\m$). We can see that the upper boundary of the parameter space coincides with the asymptotic upper bound that the top curves form. In other words, above the upper boundary of the blue region, the value of the scalar spectral index is incorrect, for any $\varphi_*$.

Increasing the range for $\varphi_{*}$ does not change the shape of the upper boundary of the parameter space in the $(\log_{10}\xi_{*},\kappa)$ plane. Indeed, we also plot more $\kappa(\xi_{*})|_{n_s=0.9649}$ curves, now in purple, with $\varphi_{*}$ ranging from $-200\,\m$ to $-30\, \m$. We find that in the range $-30\,\m$ to 0 we cover almost the entirety of the shown parameter space, while approaching more negative values simply covers a region of the parameter space discarded by observations, located at smaller and smaller values of $\kappa$.

\begin{figure}[h]
     \centering
     \begin{subfigure}[b]{0.32\textwidth}
         \centering
         \includegraphics[width=\textwidth]{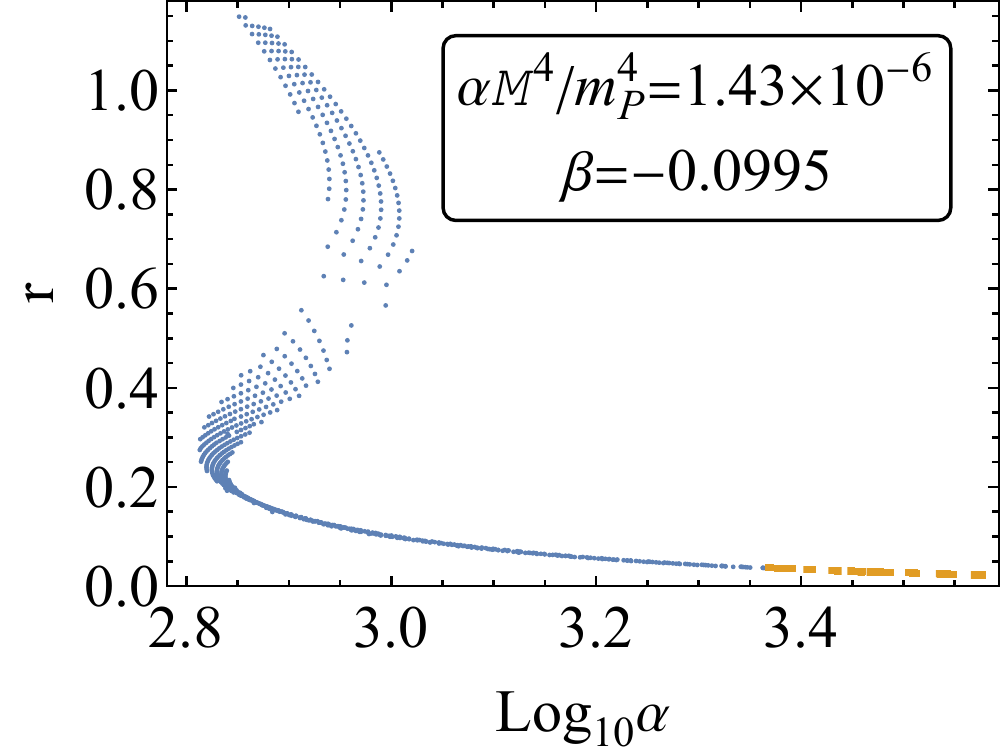}
     \end{subfigure}
     \begin{subfigure}[b]{0.32\textwidth}
         \centering
         \includegraphics[width=\textwidth]{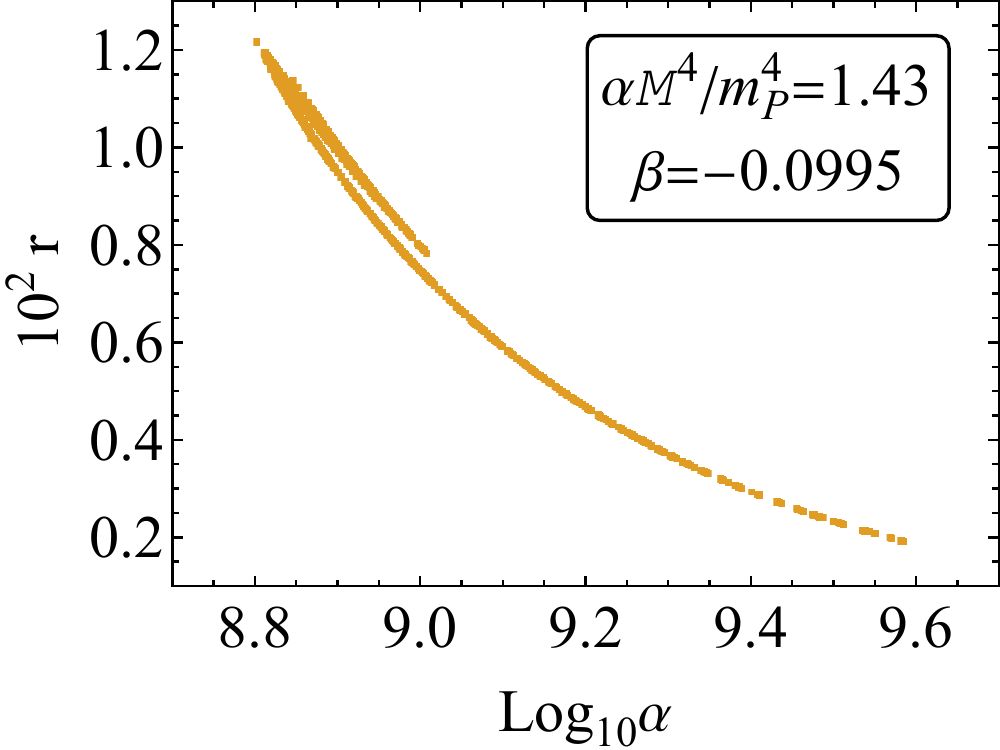}
     \end{subfigure}
          \begin{subfigure}[b]{0.32\textwidth}
         \centering
         \includegraphics[width=\textwidth]{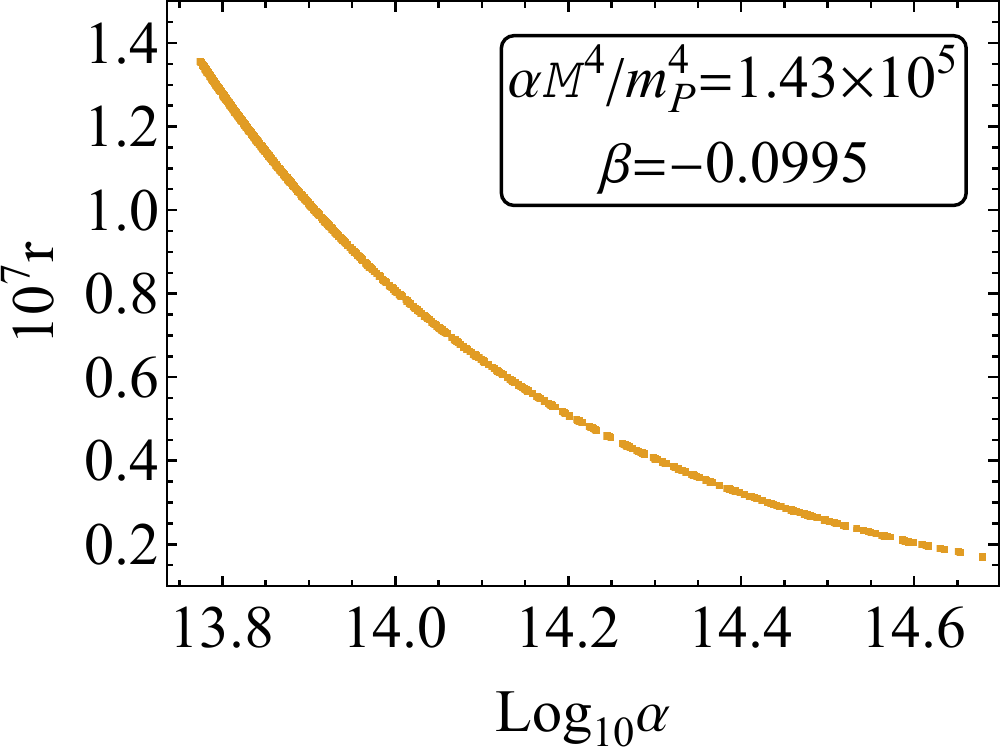}
     \end{subfigure}
     \caption{The tensor-to-scalar ratio $r$ as a function of $\log_{10}\alpha$ for different values of $\alpha M^4$, with fixed $\beta=-0.0995$. Blue points have a correct $n_s$, $\alpha_s$ and $N$ while orange points also have a correct $r$. As we make $\alpha M^4$ larger we lower the values $r$ takes. Below the threshold value of $\alpha M^4/\m^4\simeq0.143$ there still exists an orange region (left), while above it all blue points become orange (middle and right).}
      \label{fig:rasafunctionofalpha}
\end{figure}

\begin{figure}[h]
     \centering
    \begin{subfigure}[b]{0.32\textwidth}
         \centering
         \includegraphics[width=\textwidth]{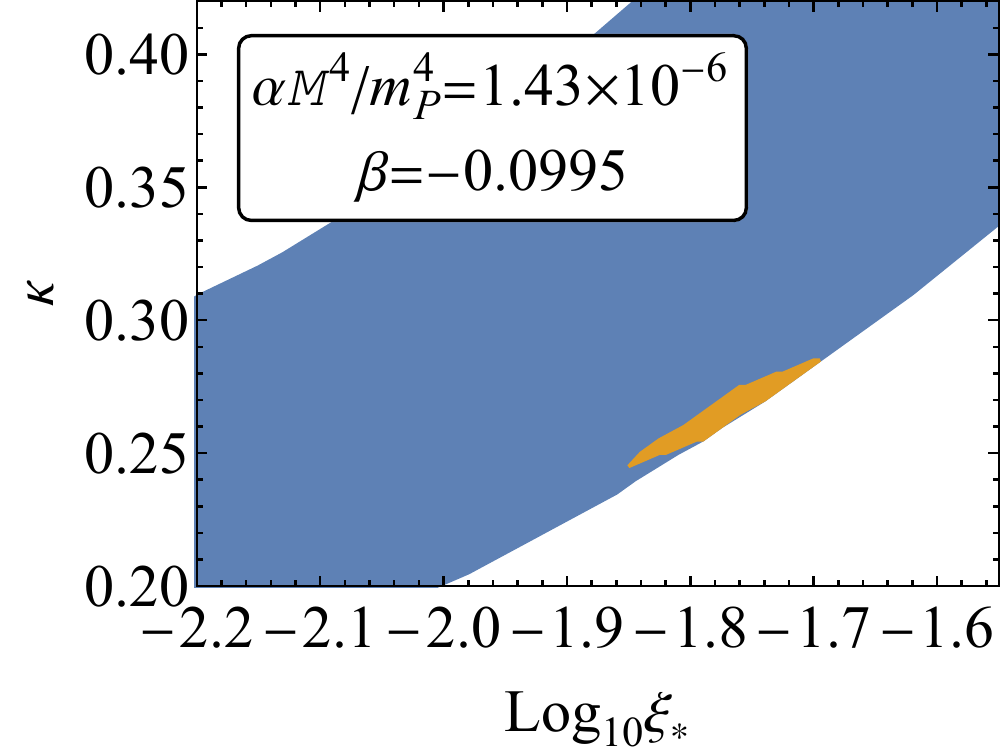}
     \end{subfigure}
     \begin{subfigure}[b]{0.32\textwidth}
         \centering
         \includegraphics[width=\textwidth]{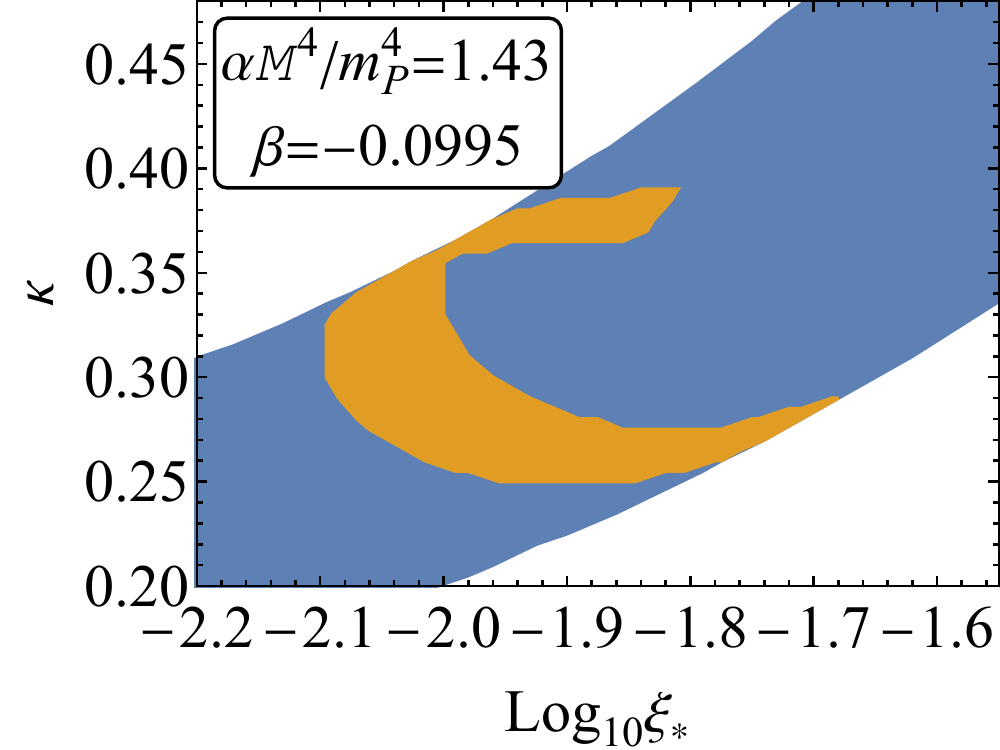}
     \end{subfigure}
     \begin{subfigure}[b]{0.32\textwidth}
         \centering
         \includegraphics[width=\textwidth]{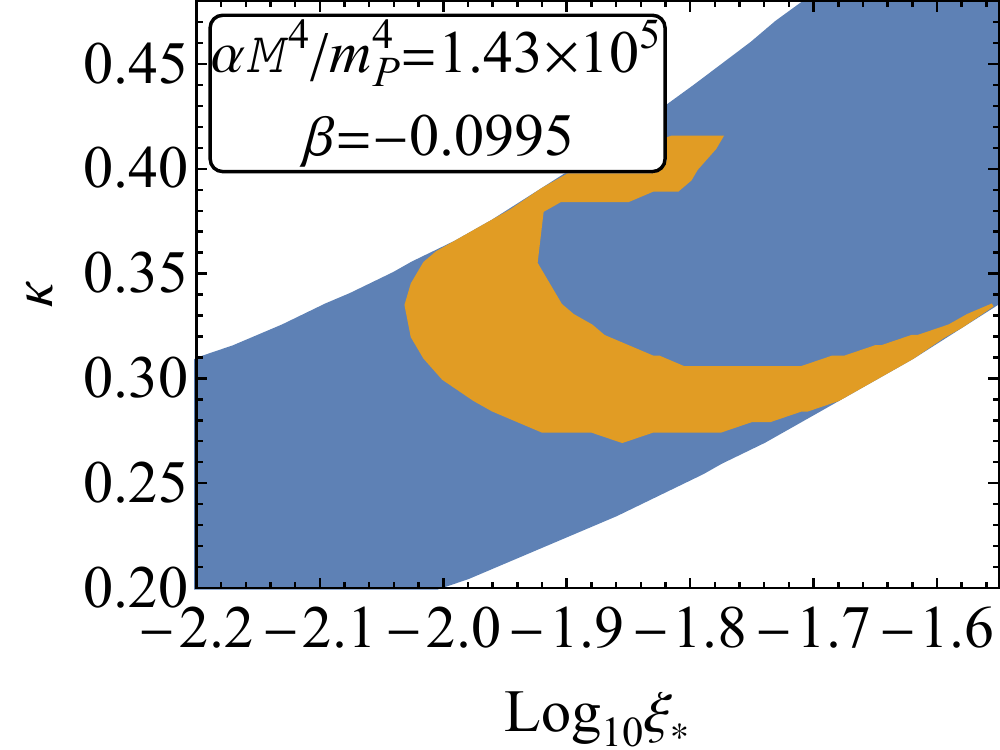}
     \end{subfigure}
     \caption{Slices of the parameter space in the $(\log_{10}\xi_{*},\kappa)$ plane with $\beta=-0.0995$ and $\alpha M^4/\m^4=1.43\times 10^{-6}$ (left), $\alpha M^4/\m^4=1.43$ (middle) and $\alpha M^4/\m^4=1.43\times10^5$ (right). The shape of the parameter space is identical for both the panels in the center and right, although the region with correct observational predictions is shifted toward larger $\kappa$ and $\xi_{*}$ as we make $\alpha M^4$ larger. Even for very small values of $\alpha M^4$ the orange region never disappears (left).}
      \label{fig:parameterspacefordifferentr}
\end{figure}

It could also be that changing $\alpha M^4$ would change the shape of the parameter space.
However, we find that the main effect of this is on $r$. Indeed, there exists a bound, given by $\alpha M^4/\m^4\simeq 0.143$ below which the size of the orange region in the parameter space is reduced in size (although it never fully disappears, see the left panel in Fig. \ref{fig:parameterspacefordifferentr}), and above which its position shifts towards larger values of $\kappa$ and $\xi$.
This can be seen by comparing the middle and right panels in Fig. \ref{fig:parameterspacefordifferentr}. It is also straightforward to see from Eqs. \eqref{eq:sr_V_in_model} and \eqref{eq:CMB_SR} that $r$ can be made arbitrarily small by making $\alpha M^4$ larger, as we have obtained in our numerical study (see Fig. \ref{fig:rasafunctionofalpha}). However, it is important to note that the shift in the orange region of the parameter towards larger $\kappa$ and $\xi$ can change the subsequent cosmological evolution after inflation ends, since these points serve as the initial conditions for the later evolution.

We conclude that the shape of the blue region shown in Fig. \ref{fig:parameterspace1} is an universal feature of the model, with the caveat that the analysis concerning Fig. \ref{fig:parameterspacecomparison} is for the $\beta=0$ case. We expect only minor modifications to this figure when studying the general non-zero $\beta$ case, since during slow-roll inflation the value of the field barely changes and we choose the scale $\mu$ to be approximately equal to the field value at horizon exit $\varphi_{*}$, making the running in $\xi$ negligible. In the same spirit, it is obvious that the shapes of the blue and orange regions in Fig. \ref{fig:parameterspacecomparison}, for which $\beta=0$, are very similar to the analogous regions in Fig. \ref{fig:parameterspace1}, for which $\beta=-0.1$.

\begin{figure}[h]
    \centering
    \includegraphics[width=0.6\textwidth]{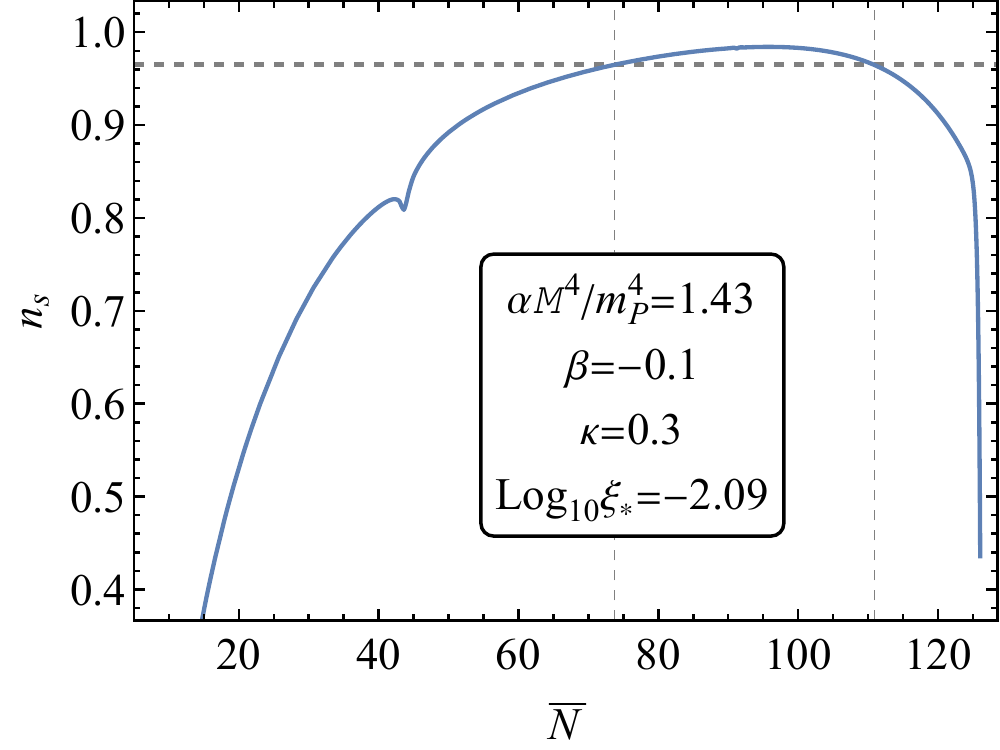}
    \caption{Scalar spectral index as a function of the number of e-folds before the end of inflation in the Einstein frame, for $\alpha M^4=1.43/\m^4$, $\beta=-0.1$, $\kappa=0.30$ and $\log_{10}\xi_{*}=-2.09$. $\bar{N}=0$ corresponds to the end of inflation. The horizontal dashed line is located at $n_s=0.9649$, and it intersects $n_s(\bar{N})$ at $\bar{N}=73.7$ and at $\bar{N}=110.8$.}
    \label{Fig:nsasafunctionofn}
\end{figure}

To conclude this section, in Fig. \ref{Fig:nsasafunctionofn}, we show an example plot of the scalar spectral index as a function of the number of e-folds before the end of inflation in the Einstein frame. The shape of $n_s(\bar{N})$ in Fig. \ref{Fig:nsasafunctionofn} is general and for most of the valid points of the parameter space, the equation $n_s(\bar{N})=0.9649$ has two solutions, \textit{e.g.}, $\bar{N}=73.7$ and $\bar{N}=110.8$ in the specific case of the figure under consideration. Of course, only one of the two is selected via the bounds imposed on the initial radiation energy density. We have not found a trend where only the first (or the second) of the solutions are the correct ones. Indeed, depending on the region of the parameter space under consideration we can have one or the other giving the correct value for the number of e-folds.

\subsection{Numerical results for post-inflationary evolution}
In order to gain some understanding about the model, we start this section by studying one specific benchmark point of the parameter space which leads to correct dark energy predictions. After this we show the full parameter space of our quintessential inflation model. 

\begin{figure}[h]
     \centering
     \begin{subfigure}[b]{0.49\textwidth}
         \centering
         \includegraphics[width=\textwidth]{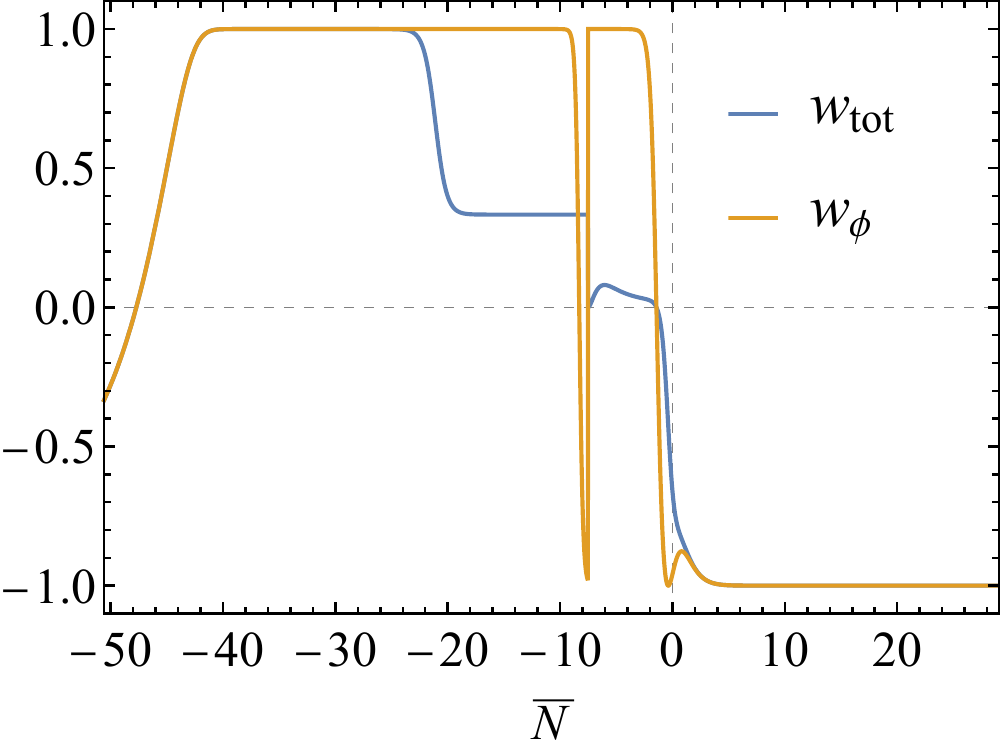}
     \end{subfigure}
     \begin{subfigure}[b]{0.49\textwidth}
         \centering
         \includegraphics[width=\textwidth]{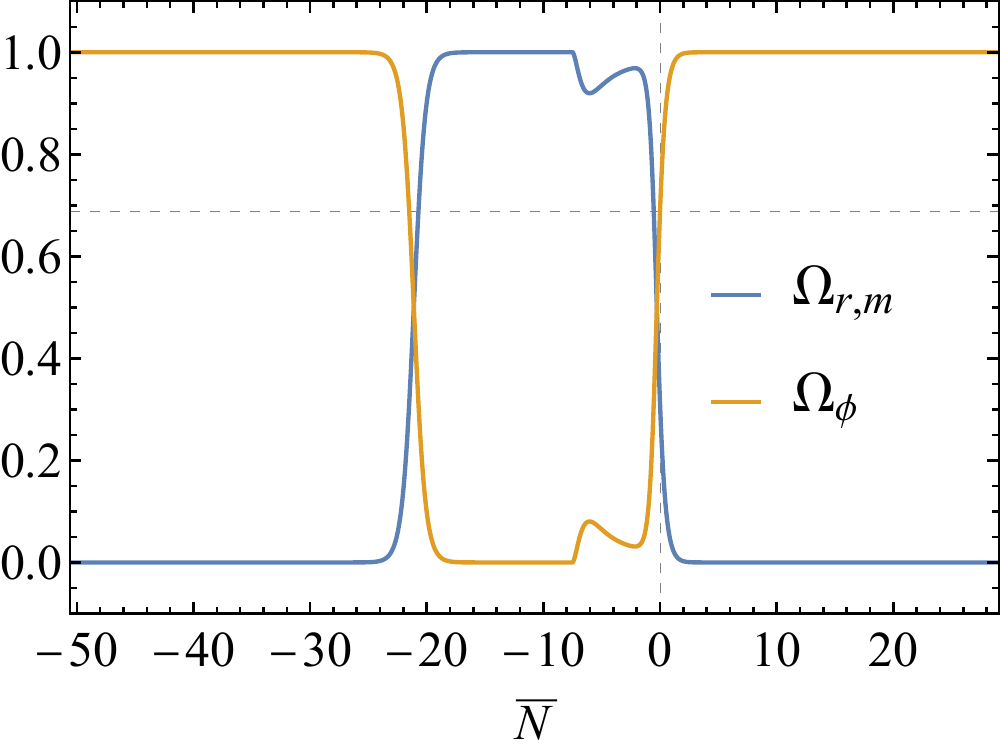}
     \end{subfigure}
     \caption{Left: Barotropic parameter of the universe (blue) and of the inflaton (orange) as a function of the elapsing number of e-folds in the Einstein frame. Right: Energy density parameter of the background fluid (blue), which is radiation (r) before and pressureless dust (m) after equality, and of the field (orange) as a function of the elapsing number of e-folds in the Einstein frame. The horizontal dashed line is located at 0.6889. For both graphs $\bar{N}=0$ corresponds to the present time and $\bar{N}=-7.5$ to matter-radiation equality.}
      \label{Fig:barotropicparameteranddensities}
\end{figure}

\begin{figure}[h]
    \centering
    \begin{subfigure}[b]{0.49\textwidth}
         \centering
         \includegraphics[width=\textwidth]{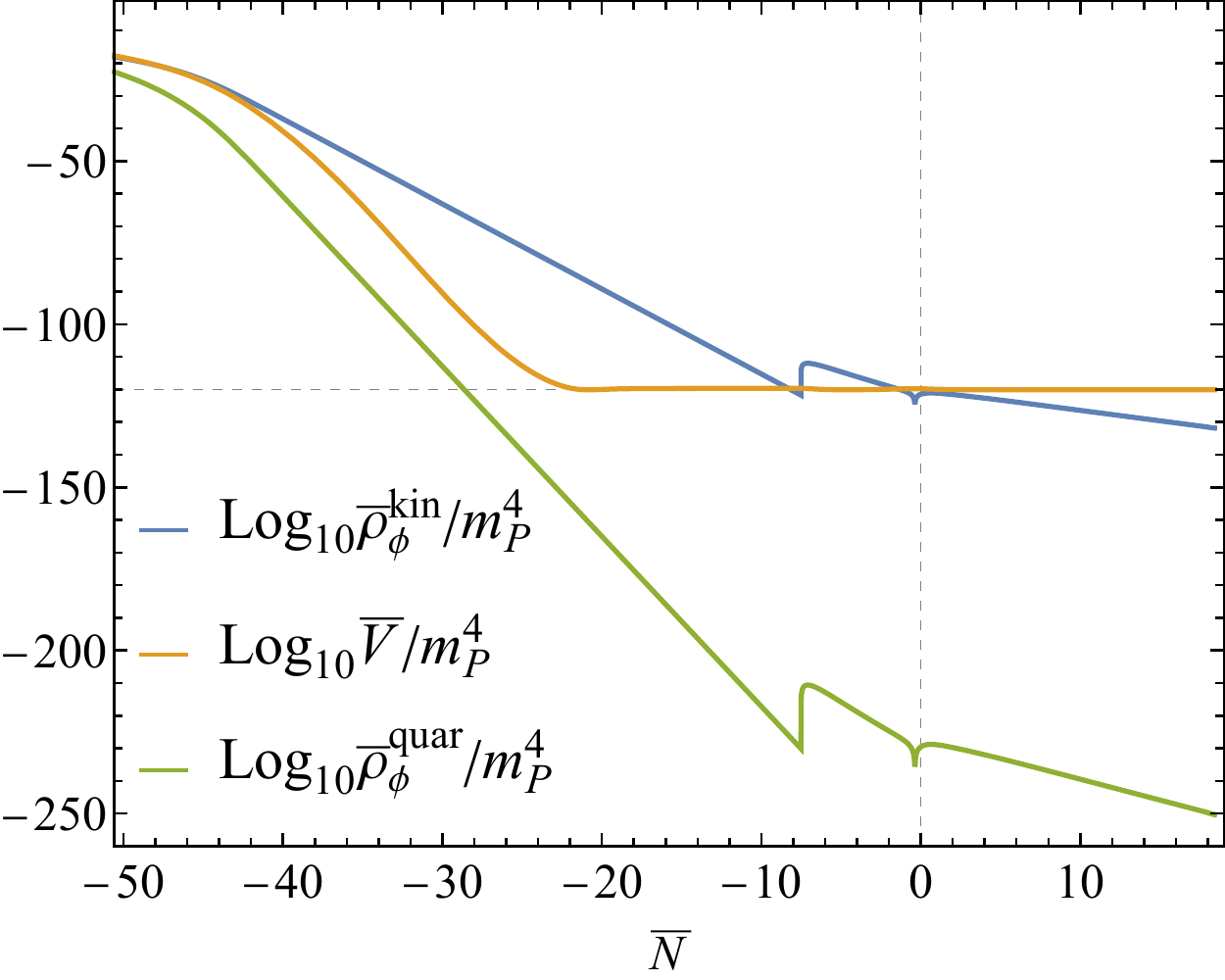}
     \end{subfigure}
     \begin{subfigure}[b]{0.49\textwidth}
         \centering
         \includegraphics[width=\textwidth]{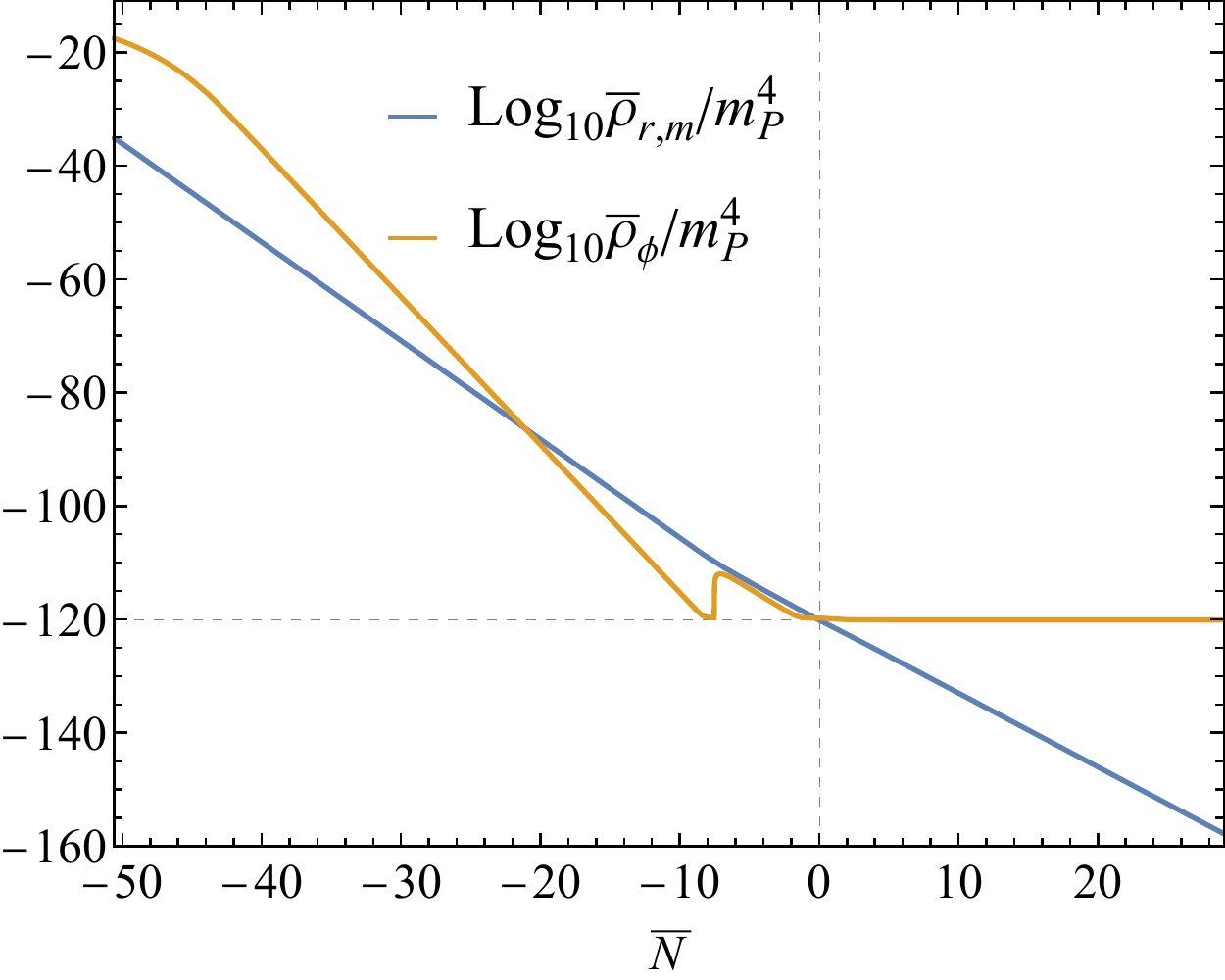}
     \end{subfigure}
    \caption{Left: Contributions from the kinetic energy energy density $\bar{\rho}_{\phi}^{\text{kin}}=\dotEforfig{\phi}^2/2$ (blue), potential energy density $\bar{V}$ (orange) and quartic kinetic term $\bar{\rho}_{\phi}^{\text{quar}}=3\alpha\qty(1+4\alpha V/h^2)\dotEforfig{\phi}^4/4$ (green) to the total energy density of the inflaton in the Einstein frame in Planck units, as a function of the elapsing number of e-folds in the Einstein frame. These contributions correspond to the first, second and third terms in the action \eqref{eq:S_Einstein}, respectively. Right: Einstein frame energy densities of the background fluid (blue), which can be either radiation (r) or pressureless dust (m), and of the inflaton (orange) as a function of the elapsing number of e-folds in the Einstein frame. The horizontal dashed lines are located at $\log_{10}(\bar{\rho}/\m^4)=-120$, $\bar{N}=0$ corresponds to the present time, and $\bar{N}=-7.5$ corresponds to matter-radiation equality.}
    \label{Fig:contributionstoenergydensity}
\end{figure}

Let us look at the point in parameter space with parameter values given by
\begin{equation} \label{eq:parameterspacepoint}
\begin{gathered}
    \kappa = 0.284 \, , \quad
    \log_{10}\xi_{*} = -1.960 \, , \quad
    \alpha = 7.73\times 10^{12} \, , \\
    M^4/\m^4 = 1.85\times 10^{-9} \, , \quad
    \beta = -0.100 \, , \quad \text{and} \quad
    \mu = -6\m \, ,
\end{gathered}
\end{equation}
which satisfies all the conditions listed above required for correct inflation and dark energy. This can be immediately confirmed by looking at Figs. \ref{Fig:barotropicparameteranddensities} and \ref{Fig:contributionstoenergydensity}. In the left panel in Fig. \ref{Fig:barotropicparameteranddensities} we show the barotropic parameter of the inflaton and of the whole universe, which are given by
\begin{equation}
    w_{\phi}=\frac{\bar{p}_{\phi}}{\bar{\rho}_{\phi}} \quad \text{and} \quad w_{\text{tot}}=\frac{w_{\text{r,m}}\bar{\rho}_{\text{r,m}}+\bar{p}_{\phi}}{\bar{\rho}_{\text{r,m}}+\bar{\rho}_{\phi}} \, ,
\end{equation}
where $w_{\text{r,m}}$ is equal to either $1/3$ or $0$ for a radiation (r) or a pressureless dust (m) background with energy density $\bar{\rho}_{\text{r,m}}$, respectively, and $\bar{\rho}_{\phi}$ and $\bar{p}_{\phi}$ are given by Eq. \eqref{eq:rho_p_Einstein}. At the present time, which corresponds to $\bar{N}=0$ in both figures, the energy fraction of the field is $\Omega_{\phi}^0=0.6889$ (see the right panel in Fig. \ref{Fig:barotropicparameteranddensities}) and its barotropic parameter and running are $w_{\phi}^0=-0.95895$ and $w_a^0=-0.17034$, in agreement with dark energy observations. As for the energy densities at present it can be confirmed by looking at the right panel in Fig. \ref{Fig:contributionstoenergydensity}, that the energy density of the field is $\bar{\rho}_{\phi}=1.7\times10^{-120}\m^4$ while that of the fluid is $\bar{\rho}_{\text{m}}=7.5\times10^{-121}\m^4$, which are within an order of magnitude of observations. Finally, the temperature of the universe at the onset of radiation domination, \textit{i.e.}, when $w_{\text{tot}}=0.36$ and $\Omega_{\phi}=0.05$, is $T=(30\rho/(\pi^2g))^{1/4}\simeq2.49\times10^{-23}\m= 0.15\,\text{MeV}$, which is slightly above $T_{\text{BBN}}$.


As a far as inflationary observables and dark energy predictions go, the point given by Eq. \eqref{eq:parameterspacepoint} is fine. However, as the careful reader might have noticed, there are two issues with the matter dominated era. As it can be seen in the left panel in Fig. \ref{Fig:barotropicparameteranddensities}, its duration $\bar{N}_{\text{mat}}=7.5$ is below what would be expected in a standard cosmology, where $\bar{N}_{\text{mat}}\sim 8$. Furthermore, the barotropic parameter of the universe is not exactly zero (although it stays below 0.1). We can explain this behaviour by taking a closer look at our model. We remind the reader that, as shown in Eqs. \eqref{eq:phi_eom_with_rho_mixing} and \eqref{eq:matter_continuity_Einstein}, there is a coupling between the inflaton and the fluid (the last term in both equations) coming from the conformal factor that appears in the matter action after the conformal transformation to the Einstein frame. During inflation we have $\bar{\rho}_{\text{r,m}}=0$ and during kination and the radiation dominated era we have that the barotropic parameter of the fluid is $w=1/3$, so that the coupling is not present until matter-radiation equality. However, as soon as we have a pressureless dust-dominated universe, with $w=0$, the coupling is turned on. In order to better understand this, after some simple algebra, one can rewrite Eq. \eqref{eq:matter_continuity_Einstein} as 
\begin{equation} \label{eq:effectivecontinuityequation}
    \dotE{\bar{\rho}} + 3 \bar{H} \bar{\rho}(1+w_{\text{eff}}) = 0 \, ,
\end{equation}
where
\begin{equation}
    w_{\text{eff}}=w+\frac{(1-3w)}{3\qty(\frac{2F_R H}{\dot{F}_R}+1)}=\frac{1}{3\qty(\frac{2F_R}{F_R'}+1)} \, ,
\end{equation}
where the last equality follows from working in the matter dominated era and a prime denotes a derivative with respect to the Jordan frame number of e-folds. Thus, $w_{\text{eff}}$ will only be close to zero when the rate of change of $F_R$ satisfies
\begin{equation} \label{eq:rateofchangeFR}
    \frac{F_R'}{F_R}\ll 1 \, .
\end{equation}
However, looking at the expression for $F_R$ in Eq. \eqref{eq:F_R}, and remembering that the terms coming from the $\alpha$ contribution are negligible at late times, the rate of change from Eq. \eqref{eq:rateofchangeFR} is approximately $F_R'/F_R\sim \varphi'/\varphi$. By noticing that the field is in free fall, and, thus, has a non-negligible rate of change, during the matter dominated era (its barotropic parameter is one\footnote{It could be that the higher order kinetic terms that appear in the Einstein frame modify the barotropic parameter of the field from its usual expression $w_\phi=(\frac{1}{2}\dotE{\phi}^2-\bar{V})/(\frac{1}{2}\dotE{\phi}^2+\bar{V})$. This is not the case, as can be seen in Fig. \ref{Fig:contributionstoenergydensity}, where it is clear that the quartic kinetic term plays a subdominant role throughout the expansion history of the universe.} as can be seen in Fig. \ref{Fig:barotropicparameteranddensities}) it immediately follows that $F_R'/F_R$ cannot be very small and $w_{\text{eff}}$ will be generally larger than zero, as we find. 

As for the number of e-folds of the matter dominated era $\bar{N}_{\text{mat}}$, noting that from Eq. \eqref{eq:effectivecontinuityequation} follows that $\bar{\rho}\propto \bar{a}^{-3(1+w_{\text{eff}})}$, a simple calculation reveals
\begin{equation}
    \bar{N}_{\text{mat}}=\log\frac{\bar{a}_0}{\bar{a}_{\text{eq}}}=\frac{1}{3(1+w_{\text{eff}})}\log\frac{\bar{\rho}_{\text{eq}}}{\bar{\rho}_0}\simeq \frac{1}{3}\log\frac{\bar{\rho}_{\text{eq}}}{\bar{\rho}_0}-\frac{w_{\text{eff}}}{3}\log\frac{\bar{\rho}_{\text{eq}}}{\bar{\rho}_0} \, ,
\end{equation}
where we have taken into account that $w_{\text{eff}}\lesssim 0.1$, as is the case for most of the valid parameter space. Thus, $\bar{N}_{\text{mat}}$ will generally be smaller than its canonical value in standard Einstein--Hilbert gravity, where there is no coupling between the fluid and the inflaton so that $w_{\text{eff}}=0$. Introducing the values of the energy density of the fluid at equality, $\bar{\rho}_{\text{eq}}=1.27\times 10^{-110}\m^4$, and at the present time, $\bar{\rho}_{0}=3.28\times10^{-121}\m^4$, we find that $\bar{N}_{\text{mat}}$ could be decreased by as much as about one e-fold.  We take this into account in the parameter space scans, not neglecting points that \textit{a priori} would have been considered to have a too short matter dominated era. In this way, we choose six as the smallest value $\bar{N}_{\text{mat}}$ can take when scanning over the parameter space, although, as we will see below, for all valid points $\bar{N}_{\text{mat}}$ will always be larger than seven, in agreement with the approximation $w_{\text{eff}}\lesssim 0.1$ that we have taken above.

\begin{figure}[h]
    \centering
    \includegraphics[width=0.6\textwidth]{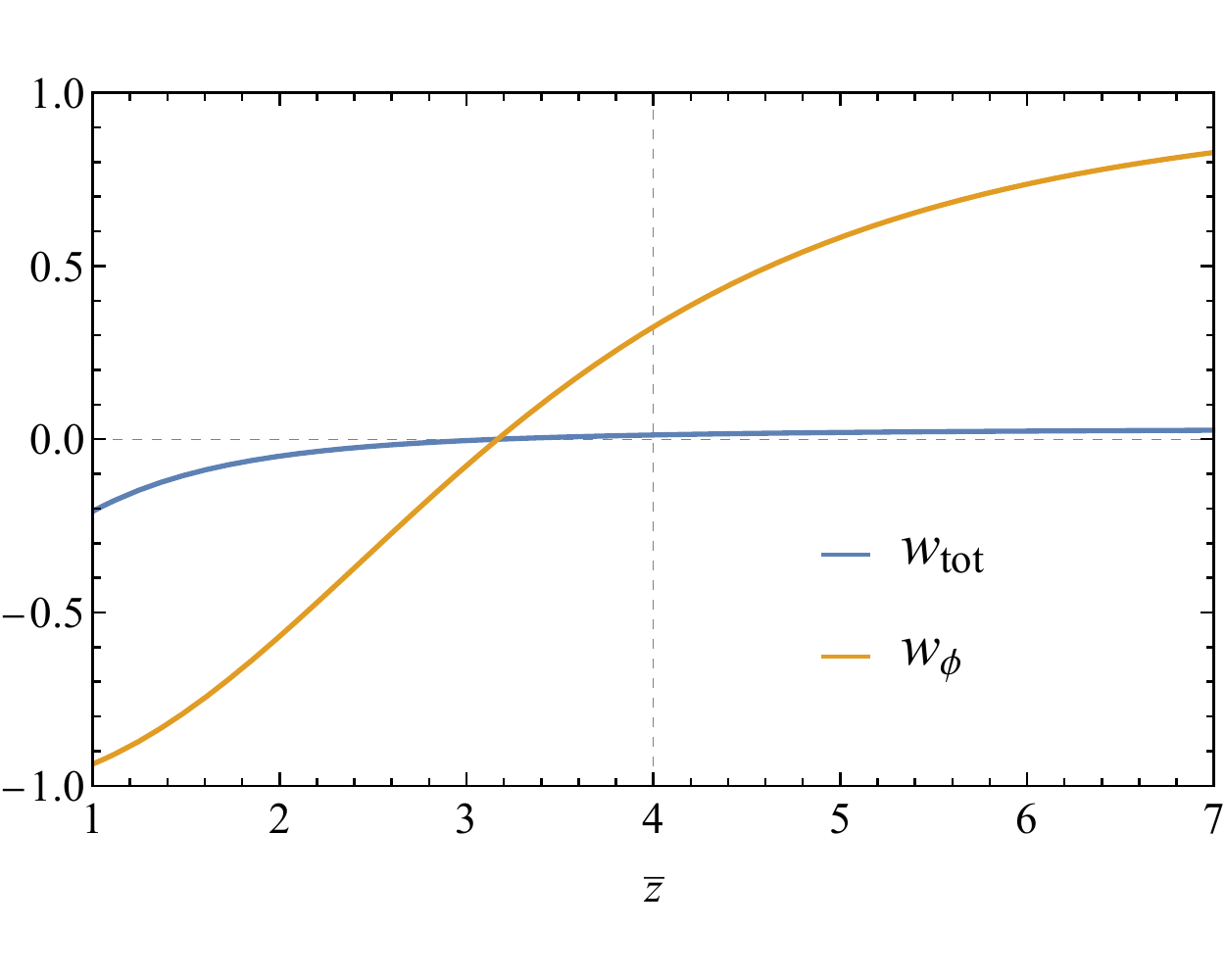}
    \caption{Barotropic parameter of the universe (blue) and of the inflaton (orange) as a function of the redshift in the Einstein frame. The vertical dashed line is located at $\bar{z}=4$, corresponding to galaxy formation. The barotropic parameter of the universe is very close to zero around this redshift, making structure formation largely unimpeded.}
    \label{Fig:barotropicparameterasafunctionofredshift}
\end{figure}

In conclusion, we have obtained that the barotropic parameter of the universe during the matter dominated era will generally be larger than zero and that the length of this era will generally be shorter than in Einstein--Hilbert gravity. These effects are an inevitable consequence of working in our modified gravity setup. However, we find that for most of the parameter space $w_{\text{eff}}\lesssim 0.1$ (and discard the points which do not satisfy this), and in fact, around redshifts corresponding to galaxy formation, \textit{i.e.}, $\bar{z}\sim 4$ (where $\bar{z}\equiv \bar{a}^{-1} - 1$) \citep{Thomas:2016hms}, the barotropic parameter is very close to zero, thereby not significantly impeding structure formation (see Fig. \ref{Fig:barotropicparameterasafunctionofredshift}).

Having discussed the effect of the inflaton-fluid coupling, modified gravity manifests itself in the Einstein frame through one other effect: the existence of a quartic kinetic term in the action (see the third term in Eq. \eqref{eq:S_Einstein}), which \textit{a priori} cannot be discarded. However, as it can be seen from the left panel in Fig. \ref{Fig:contributionstoenergydensity}, it remains subdominant throughout the expansion history of the universe. This is a general behaviour in all the valid parameter space. In what follows we neglect this term.

\begin{figure}[h]
    \centering
    \includegraphics[width=0.6\textwidth]{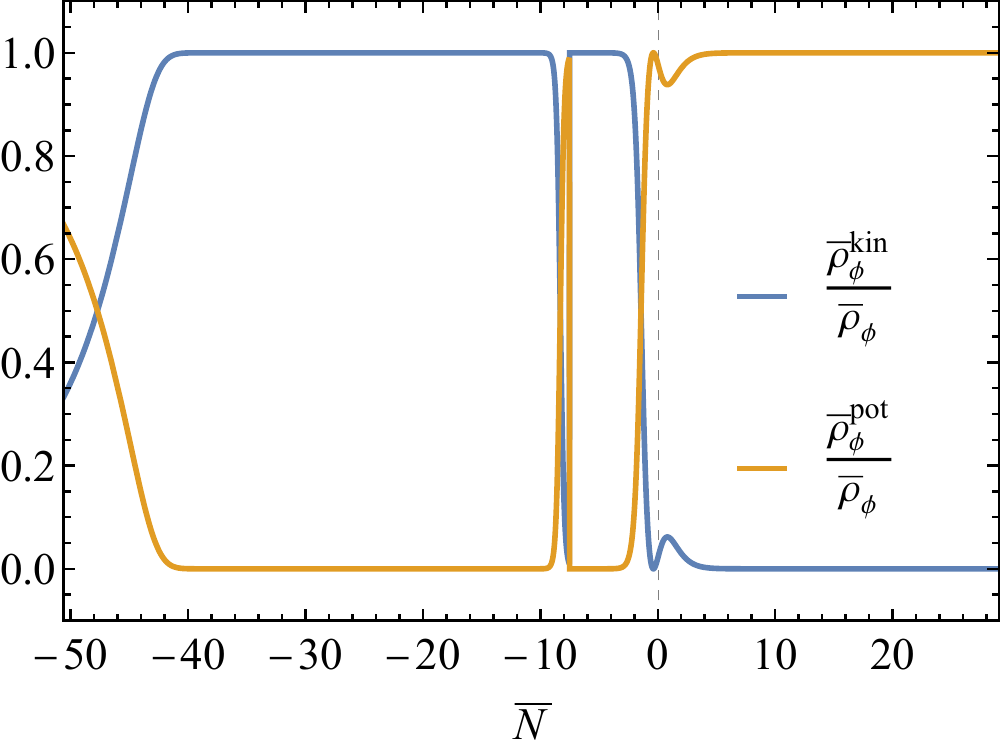}
    \caption{Kinetic energy density of the field over its total energy density (blue) and potential energy density of the field over its total energy density (orange) in the Einstein frame as a function of $\bar{N}$, from the end of inflation, at $\bar{N}=-50.6$ to the present time, at $\bar{N}=0$. The end of kination (reheating) occurs at $\bar{N}=-19.6$ and matter-radiation equality at $\bar{N}=-7.5$.}
    \label{Fig:ratios}
\end{figure}

Let us next examine the evolution of the system more carefully, stage by stage. As the field approaches the end of the inflationary plateau and its velocity starts increasing, the condition $\epsilon_H=1$ is satisfied and inflation ends. After the end of inflation there is a transition period where the field is gaining kinetic energy although its total energy density is still not dominated by it. This can be seen from Fig. \ref{Fig:ratios}, where we show the energy density ratios
\begin{equation}
    \frac{\bar{\rho}_{\phi}^\text{kin}}{\bar{\rho}_{\phi}}=\frac{\frac{1}{2}\dotE{\phi}^2}{\frac{1}{2}\dotE{\phi}^2+\bar{V}} \quad \text{and} \quad \frac{\bar{\rho}_{\phi}^\text{pot}}{\bar{\rho}_{\phi}}=\frac{\bar{V}}{\frac{1}{2}\dotE{\phi}^2+\bar{V}} \, .
\end{equation}
Indeed, after the end of inflation, at $\bar{N}=-50.6$, it is not until $\bar{N}\sim -40$ that the energy density of the inflaton is kinetically dominated, while the energy density of the universe is still dominated by that of the field ($\Omega_{\phi}$ is still equal to one as can be seen from the right panel in Fig. \ref{Fig:barotropicparameteranddensities}), giving way to the kination era. This can also be seen from the left panel in Fig. \ref{Fig:barotropicparameteranddensities}, where the barotropic parameter of the field does not become equal to one until $\bar{N}\sim -40$. Of course, at the moment when the field becomes kinetically dominated, remembering the quartic kinetic terms are negligible, we have
\begin{equation}
    w_\phi=\frac{\frac{1}{2}\dotE{\phi}^2-\bar{V}}{\frac{1}{2}\dotE{\phi}^2+\bar{V}}=\frac{\frac{1}{2}\dotE{\phi}^2}{\frac{1}{2}\dotE{\phi}^2}=1 \, .
\end{equation}
During kination, the radiation energy density fraction approaches that of the field, until it takes over and approaches one around $\bar{N}=-19.6$, see Fig.~\ref{Fig:barotropicparameteranddensities}. This moment corresponds to reheating. It is important to note that the scaling of the energy density of the universe between $\bar{N}=-50.6$ and $\bar{N}=-40$ is not $\bar{\rho}\propto \bar{a}^{-6}$ (but slower), and thus the exponent 6 on the right-hand-side of Eq. \eqref{rhoreh} is only an approximation. Effectively this means that we are able to have a reheating temperature close to $T_{\text{BBN}}$ without violating the bound on $\Omega_{_{\text{GW}}}^{\text{end}}$ discussed in Sec. \ref{sec:reheating}.

After reheating, the universe is dominated by the background radiation, while the field is still in free-fall, with its energy density being kinetically dominated. This can be seen from the left panel in Fig. \ref{Fig:barotropicparameteranddensities}, where the barotropic parameter of the field is still equal to one, as well as from the left panel in Fig. \ref{Fig:contributionstoenergydensity} and from Fig. \ref{Fig:ratios}. This behaviour continues until briefly before matter-radiation equality, when the field runs out of kinetic energy and starts to freeze (see Eq. \eqref{eq:delta_phi}). Indeed, its barotropic parameter approaches minus one (this can also be seen from Fig. \ref{Fig:ratios}, where the kinetic density ratio goes from one to zero and vice versa for the potential density ratio). However, the field never fully freezes. This is due to the change in the barotropic parameter of the background from $1/3$ to $0$ at matter-radiation equality. As explained above, at this point the coupling between the field and the fluid is turned on and there is an energy transfer between the components. One way to understand this is by noting that $w_{\text{eff}}$ is larger than zero, meaning that the background dilutes faster than in the canonical case, feeding its energy into the kinetic energy of the field. Indeed, the barotropic parameter of the field jumps back to unity and the inflaton goes back into free-fall during the entirety of the matter-dominated era, only to run out of kinetic energy and freeze again (its barotropic parameter going back to minus one) at the end of it.
\begin{figure}[h]
     \centering
    \begin{subfigure}[b]{0.48\textwidth}
         \centering
         \includegraphics[width=\textwidth]{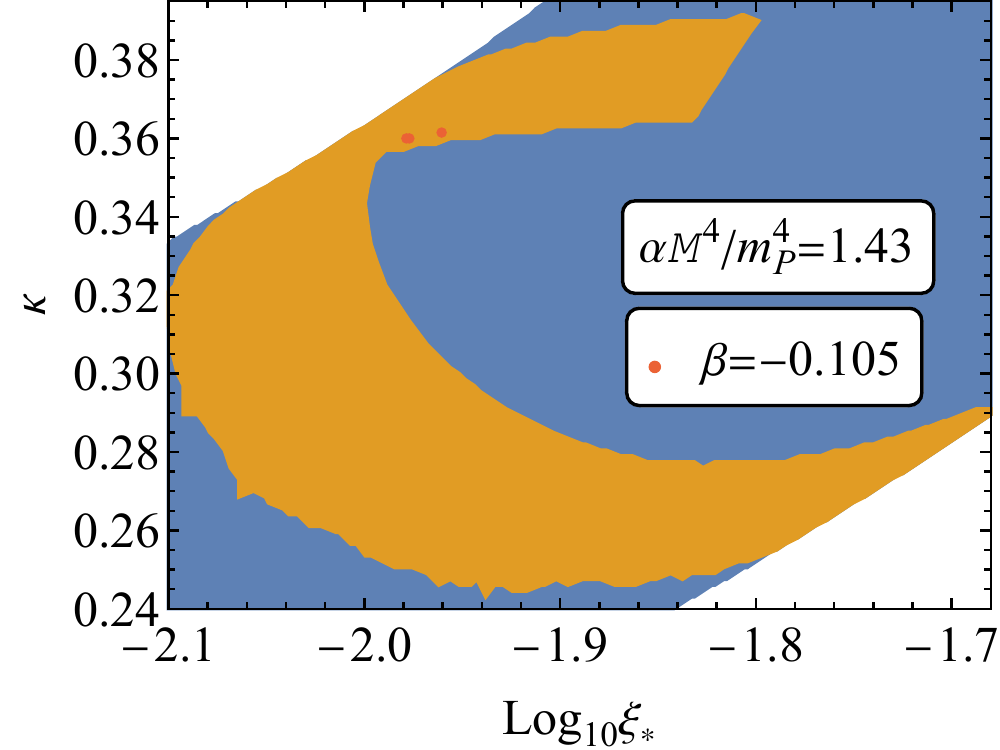}
     \end{subfigure}
     \begin{subfigure}[b]{0.48\textwidth}
         \centering
         \includegraphics[width=\textwidth]{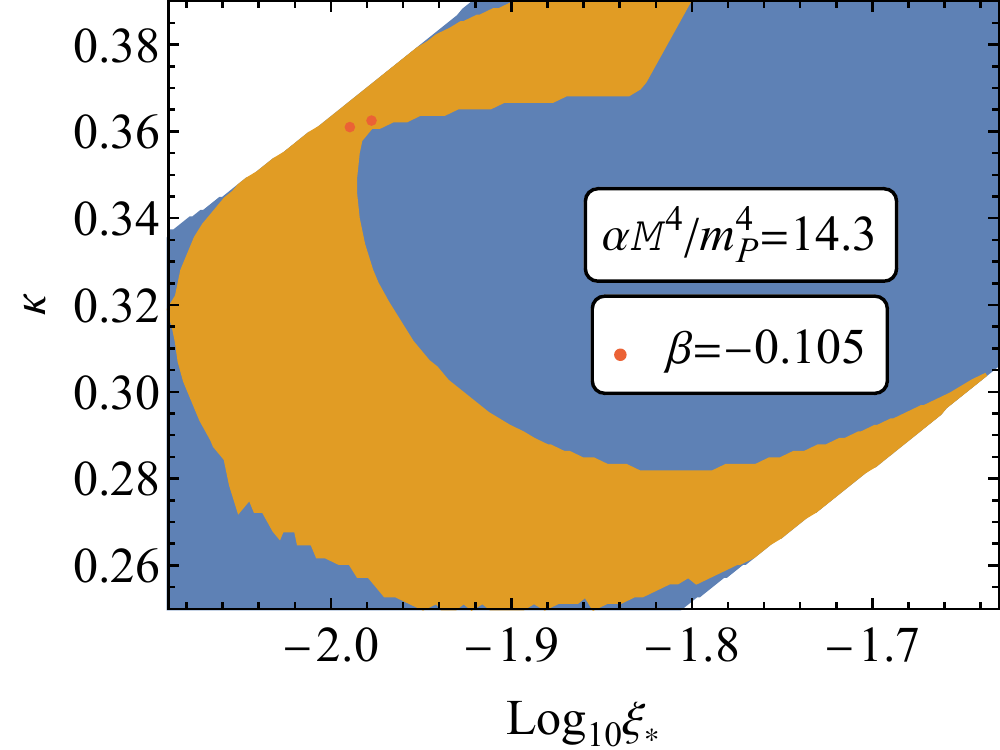}
     \end{subfigure}
     
     \bigskip
     \begin{subfigure}[b]{0.48\textwidth}
         \centering
         \includegraphics[width=\textwidth]{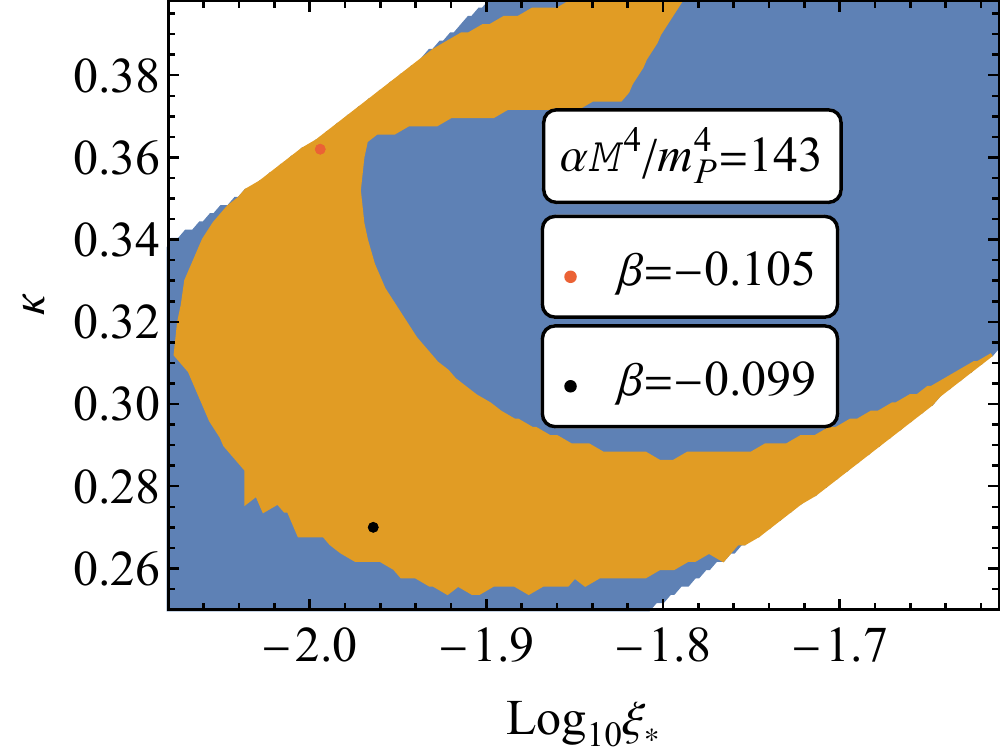}
     \end{subfigure}
     \begin{subfigure}[b]{0.48\textwidth}
         \centering
         \includegraphics[width=\textwidth]{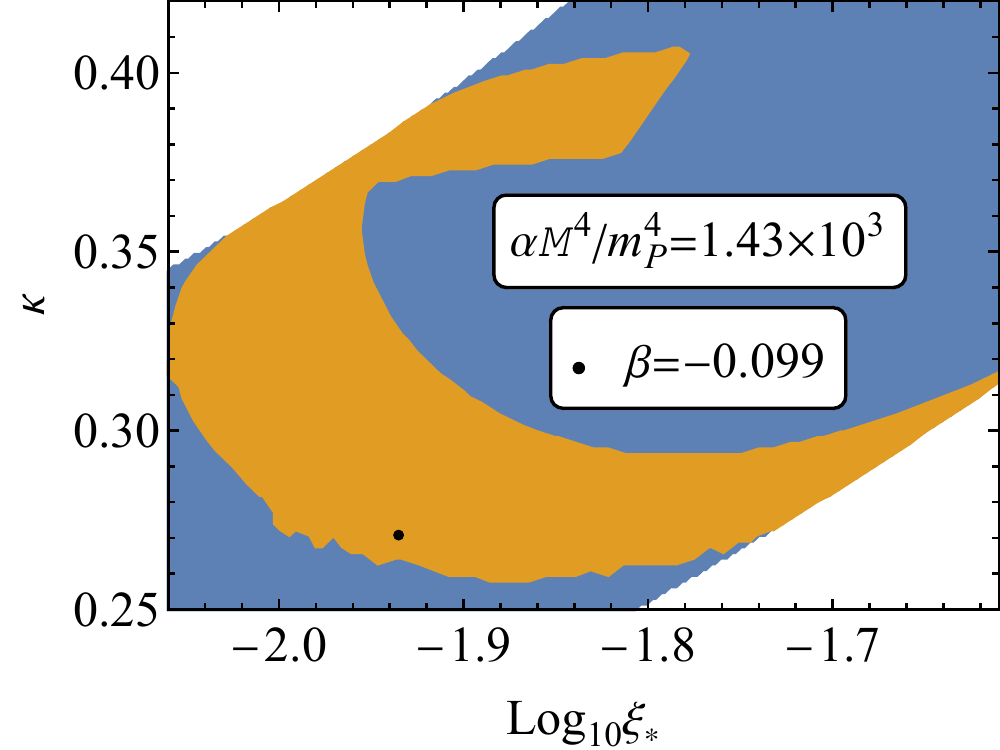}
     \end{subfigure}
     \caption{Slices of the parameter space in the $(\log_{10}\xi_{*},\kappa)$ $\alpha M^4/\m^4=1.43$ (up left), $\alpha M^4/\m^4=14.3$ (up right), $\alpha M^4/\m^4=143$ (down left) and $\alpha M^4/\m^4=1.43\times 10^3$ (down right).  Points in the blue region have a correct value of $n_s$, while points in the orange region satisfy the whole set of constraints for inflation. Red points satisfy the constraints for dark energy, while black points also satisfy strongest the bound on $\Omega_{_{\text{GW}}}^{\text{end}}$ coming from BBN 
     (where $\hat{\xi}\sim\mathcal{O}(1)$
    ). In the blue and orange regions $\beta$ takes values from the interval $[-0.108,-0.099]$ in steps of $10^{-3}$, while points giving rise to correct dark energy are only found when either $\beta=-0.099$ or $\beta=-0.105$.}
      \label{fig:deparameterspace}
\end{figure}
Finally, the field does not simply slow down and freeze. If it did, we would not find the small bump in its barotropic parameter after $\bar{N}=0$ in Fig. \ref{Fig:barotropicparameteranddensities}. The same bump can be found in Fig. \ref{Fig:ratios}. This is due to the local minimum of the potential in the Einstein frame, located slightly before the local maximum around $1+\xi(\varphi_\text{max})\varphi_{\text{max}}^2/\m^2=0$ (see discussion in Sec. \ref{sec:model}). Indeed, the field overshoots the minimum and gains some kinetic energy, only to fall back to the minimum at $\varphi_{\text{min}}=884.03\m$ and finally freeze. The present time $\bar{N}=0$ corresponds to some time briefly after overshooting the minimum but before turning back.

Having characterised the dynamics of a typical valid parameter space point, including the effects of the modified gravity terms, let us now turn our attention to the location and shape of the full valid parameter space. We show some example slices in the $(\log_{10}\xi_{*},\kappa)$ plane for different values of $\alpha M^4$ in Fig.~\ref{fig:deparameterspace}. We also scan over the parameter $\beta$, which in the orange and blue regions in the figure takes values in the interval $[-0.11,-0.098]$ in steps of $10^{-3}$. We find that points giving rise to correct dark energy (shown in red and black), which satisfy the whole set of constraints given above, are only found for $\beta=-0.099$ and $\beta=-0.105$. We also show in Tab.~\ref{table} the actual parameter values all of the successful points take. 
We find they form no specific shape in the $(\log_{10}\xi_{*},\kappa)$ plane, but expect a higher-resolution scan to reveal more working points. Lowering the required minimum temperature of the universe at the onset of the radiation dominated era, such that it is no longer larger than $T_{\text{BBN}}$, makes the valid parameter space follow a curved area inside the orange region. However, imposing the appropriate bound spoils this behaviour. It is worth mentioning that although our selection criteria regarding the length of the matter dominated era is for it to be longer than 6 e-folds, allowing for a non-zero $w_{\text{eff}}$ to decrease $\bar{N}_{\text{mat}}$, all valid points actually have at least 7 e-folds, although they are always below 8 e-folds. It is possible that the rest of constraints regarding the energy density and the barotropic parameter make the parameter space to lie in this interval.

To conclude, in this section we have characterised the behaviour, both for the field dynamics and for the modified gravity effects, of a typical successful point in the parameter space. We have also found the location of the valid points in the $(\log_{10}\xi_{*},\kappa)$ plane, having scanned over $\beta$ in the $[-0.11,-0.098]$ interval in steps of $10^{-3}$ and over $\alpha M^4/\m^4$ in the interval $[1.43,1.43\times10^3]$ in steps of factor 10. We obtain definite predictions for all of the parameters of our model except for $\alpha M^4$, which just needs to be larger than a given lower bound $\alpha M^4\sim 0.1$\footnote{This can be understood by remembering that $r=16\epsilon_V$, with $\epsilon_V$ given by Eq. \eqref{eq:sr_V_in_model}, and noting how $\alpha M^4$ enters the denominator, so that $r$ can be made arbitrarily small by making $\alpha M^4$ arbitrarily large. }. Indeed, the most  
successful points have $\kappa=0.27$, $\log_{10}\xi_{*}=-1.9$ and $\beta=-0.099$.

\begin{table}
\centering
\begin{tabular}{ |p{0.2cm}||p{0.7cm}|p{1.1cm}|p{1.5cm}|p{1.2cm}|p{1.8cm}|p{1.8cm}|p{1.2cm}|p{1.2cm}|p{0.7cm}|  }
 \hline
 \multicolumn{10}{|c|}{Successful Parameter Space Points} \\
 \hline
  & $\kappa$ & $\log_{10}\xi_{*}$ &  $\alpha M^4/\m^4$  & $\beta$& $H_{\text{CMB}}/\m$ & $10^{120}\bar{\rho}_0/\m^4$ & $w_{\phi}^0$&$w_a^0$&$\hat{\xi}$\\
 \hline
1 & 0.36 & -1.96 & 1.43 & -0.105& $9.22\times10^{-6}$&$2.31$ & -0.954 & -0.185 & 0.36\\
2 & 0.36 & -1.98 & 1.43 & -0.105& $9.17\times10^{-6}$ & $6.69$& -0.960 & -0.168 & 0.48\\
3 & 0.36 & -1.97 & 1.43 & -0.105& $9.17\times10^{-6}$&$6.05$ & -0.969 & -0.149 & 0.39\\
4 & 0.36 & -1.98 & 14.3 & -0.105& $2.95\times10^{-6}$&$1.26$ & -0.955 & -0.179 & 0.35\\
5 & 0.36 & -1.99 & 14.3 & -0.105& $2.96\times10^{-6}$&$3.66$ & -0.959 & -0.168 & 0.45\\
\textbf{6} & \textbf{0.27} & \textbf{-1.96} & \textbf{143}  & \textbf{-0.099}& \textbf{$9.78\times10^{-7}$}&\textbf{$2.44$} & \textbf{-0.951} & \textbf{-0.184} &\textbf{1.01}\\
7 & 0.36 & -1.99 & 143 & -0.105&$9.56\times10^{-7}$ &$1.92$ & -0.960 & -0.166 & 0.39\\
\textbf{8} & \textbf{0.27} & \textbf{-1.92} & \textbf{1430} & \textbf{-0.099}& \textbf{$2.89\times10^{-7}$}&\textbf{$3.63$} & \textbf{-0.951} & \textbf{-0.194} &\textbf{0.90}\\
 \hline
\end{tabular}
\caption{Parameter values for the parameter space points which give rise to successful inflation and dark energy. For each point we also show the value of $\hat{\xi}$, the Hubble parameter at the time at which the cosmological scales exit the horizon (in Planck units), energy density of the universe (in Planck units), the barotropic parameter of the field and its running, all at the present cosmic time. The two points which satisfy strongest the lower bound on $\hat{\xi}$ are highlighted in bold.}
\label{table}
\end{table}

\section{Conclusions}
\label{sec:conclusions}

In this paper we studied a relatively simple model of quintessential inflation where a single scalar field can unify the two epochs of accelerated expansion in the history of the Universe: inflation and dark energy domination. We worked in the framework of Palatini gravity where the metric and the connection are treated as independent variables. The three main ingredients in our action are:
\begin{itemize}
    \item An exponential potential of the form $M^4e^{-\kappa\varphi/\m}$ which for large positive values of the scalar field produces the quintessential tail.
    \item An $\alpha R^2$ term which asymptotically flattens the potential for large negative values and produces inflation in agreement with observations.
    \item A non-minimal coupling $\xi \varphi^2 R$ between the quintessence/inflaton field and gravity, where $\xi\approx\xi_*$ is approximately constant and positive during inflation but then runs to negative values with a slope $\beta$ in order to reproduce the correct late-time dark energy. Note that the region where $\xi(\varphi)$ is negative is never probed since the field freezes before that.
\end{itemize}
The main advantage of employing the Palatini formalism is that the auxiliary field introduced in order to parametrise the $R^2$ term turns out to be non-dynamical and can therefore be eliminated through its equation of motion. The resulting action is then single field, but contains a quartic kinetic term and a modified effective potential. For sufficiently large values of $\alpha$, the effective potential is always asymptotically flat and can therefore accommodate slow-roll inflation.

In addition to the quintessence/inflaton field, we considered an ideal fluid representing the matter and radiation content of the universe. We began our analysis by examining the equations of motion for the field and the fluid in both the Jordan and Einstein frames, while at the same time relating the quantities of interest in the two frames. We determined the Jordan frame equations to be easier to solve numerically. We then studied separately all the phases arising during the time evolution of our model in a cosmological setup, namely, inflation, kination, reheating, radiation and matter domination, and finally quintessence. To produce the radiation component after inflation, we considered as an example Ricci reheating \cite{Dimopoulos:2018wfg,Opferkuch:2019zbd,Bettoni:2021zhq}, where an additional scalar field with a non-minimal coupling to gravity reheats the universe during a period of kination. For quintessence, we showed that the Einstein frame scalar field potential develops a local minimum where the field eventually gets stuck, behaving like dark energy afterwards. The minimum is generated by the non-minimal coupling of the scalar field running to negative values. The dark energy density there is generated through the interplay of the different parameters, all taking natural values, avoiding the extreme fine-tuning of the cosmological constant in the standard $\Lambda$CDM scenario.

In the end, we presented a thorough analysis of our numerical procedure and results.
We scanned over the inflationary parameter space and showed that, for correct choices of the parameter values, the inflationary predictions of the model match the Planck observations \cite{Planck:2018jri}. For late-time evolution, we noted the emergence of a coupling between the fluid and the scalar field, present in the Einstein frame during matter domination. This coupling turned out to be the biggest obstacle for our model building, threatening to disrupt the standard cosmic evolution by transferring energy from the matter fluid to the rolling field. Nevertheless, we found example points that satisfy all the criteria we set for a successful cosmological scenario, in particular for the present-time energy density and barotropic parameter of the quintessential dark energy component. We obtain definite predictions for all of the parameters of our model. The preferred parameter values which give rise to 
successful results are  $\kappa=0.27$, $\log_{10}\xi_{*}=-1.9$ and $\beta=-0.099$. We did not find a preference for any specific value for the combination $\alpha M^4$, as long as it is above the threshold $\alpha M^4/m_{\rm P}^4\sim0.1$, below which the tensor-to-scalar ratio is too large to be compatible with observations. In addition to satisfying all the available observational constraints, our model also offers testable predictions, to be probed in the future by experiments such as EUCLID \cite{Euclid:2021qvm}. Indeed, a non-zero derivative of the barotropic parameter of dark energy with respect to the scale factor ($w_a$ in the CPL parametrization), as is the case in our model, would favor dynamical dark energy models over a cosmological constant (as in $\Lambda$CDM). Our model offers specific predictions for $w_a$, which will be useful to discern between dynamical dark energy models as measurements become more precise. It also features a non-zero barotropic parameter of the universe, probing redshifts between galaxy formation and equality, \textit{i.e.}, $\bar{z}\sim$500--1500.

To conclude, our model produces successful inflation and quintessential dark energy from the above-listed simple set of ingredients alone, without the extreme fine-tuning of $\Lambda$CDM. Our model is the first one 
(barring the toy-model in Ref.~\cite{Dimopoulos:2020pas})
to produce successful quintessential inflation using modified gravity as the main ingredient.

\acknowledgments

KD is supported (in part) by the Lancaster-Manchester-Sheffield Consortium
for Fundamental Physics under STFC grant: ST/T001038/1.
AK was supported by the Estonian Research Council grants PSG761, MOBTT5, MOBTT86, and by the EU through the European Regional Development Fund CoE program TK133 ``The Dark Side of the Universe". SSL is supported by the FST of
Lancaster University. ET was supported by the Estonian Research Council grants PRG803, MOBTT5, and PRG1055 and by the EU through the European Regional Development Fund CoE program
TK133 ``The Dark Side of the Universe".

\appendix

\section{Solving for the Hubble parameter}
\label{sec:solving_H}

In this appendix, we solve the Jordan frame Hubble parameter $H$ in \eqref{eq:Palatini_Friedmann} explicitly in terms of $\varphi$, $\dot{\varphi}$, and the fluid energy density $\rho$. We begin by using \eqref{eq:F_R}, \eqref{eq:matter_continuity_Jordan}, and \eqref{eq:phi_eom_Jordan} to decompose the time derivative of $F_R$ as
\begin{equation}
    \partial_0 F_R=A+H B \, ,
    \label{aodifasndf}
\end{equation}
where
\begin{eqnarray}
    A=&&\frac{2\varphi\dot{\varphi}\tilde{\xi}}{\m^2}+\frac{2\alpha\dot{\varphi}\left(3V'(\varphi)-\tilde{\xi}\varphi R\right)}{\m^4\left(1+\frac{\xi\varphi^2}{\m^2}\right)}+\frac{2\alpha\varphi\dot{\varphi}\tilde{\xi}}{\m^6\left(1+\frac{\xi\varphi^2}{\m^2}\right)^2}T
    \label{adbfbahsdf}
\end{eqnarray}
and
\begin{equation}
    B=\frac{3\alpha}{\m^4} \left[\frac{2\dot{\varphi}^2-\rho(1+w)(1-3w)}{1+\frac{\xi\varphi^2}{\m^2}}\right] \, ,
    \label{adifbauhsdfasdf}
\end{equation}
where $\tilde{\xi}$ and $T$ are given by Eqs. \eqref{eq:xitilde} and \eqref{eq:energymomentumtrace} respectively. Note that by using Eq. \eqref{eq:R}, the expression for $A$ can be simplified further to obtain
\begin{equation}
    A=\frac{2\varphi\dot{\varphi}\tilde{\xi}}{\m^2}\left(1+\frac{2\alpha T}{\m^4\left(1+\frac{\xi\varphi^2}{\m^2}\right)^2}\right)+\frac{6\alpha\dot{\varphi}V'}{\m^4\left(1+\frac{\xi\varphi^2}{\m^2}\right)}.
\end{equation}
With these, Eq. \eqref{eq:Palatini_Friedmann} can be recast as 
\begin{equation}
    H^2\qty(3F_R+3B+\frac{3}{4F_R}B^2)+H\qty(3A+\frac{3}{2F_R}AB)+\frac{3}{4F_R}A^2-\frac{T_{00}}{\m^2}-\frac{\alpha}{4\m^2}R^2=0
    \label{adofuasujdbfn}
\end{equation}
and solved for $H$ as
\begin{equation}
    H=-\frac{A}{B+2F_R}+\frac{\sqrt{3F_R(4T_{00}+\alpha R^2)}}{3(B+2F_R)} \, ,
\end{equation}
with $R$, $F_R$, and $T_{00}$ from Eqs.~\eqref{eq:R}, \eqref{eq:F_R} and \eqref{eq:T00}, respectively.
Demanding $H$ to be real sets the requirement $F_R\geq 0$ (note that $T_{00}$ and $R^2$ are always positive). From Eq. \eqref{eq:F_R}, this reads
\begin{equation}
    \qty(1+\frac{\xi\varphi^2}{\m^2})^2>\frac{\alpha T}{\m^4}=\frac{\alpha}{\m^4}\qty(\dot{\varphi}^2-4V(\varphi)-\rho(1-3w)) \, .
\end{equation}
During inflation, the background matter energy density $\rho=0$, and the condition is always satisfied when the potential dominates the kinetic term, $4V(\varphi)>\dot{\varphi}^2$, in particular during slow-roll. It is also easy to satisfy later on, when $\rho>0$ becomes important and $\alpha$ contributions become irrelevant.

\section{A bound on the bare mass-squared of the spectator field}
\label{sec:spectator_mass}

Let us estimate the upper bound of $|m^2|$, the mass squared of the spectator field $\psi$ from \eqref{Vpsi}, such that it remains
negligible at least until reheating. Firstly, let us obtain an upper bound of
the value of $\langle\psi^2\rangle$ at the end of inflation. Imposing the
requirement that \mbox{$\bar{\rho}_\psi^{\rm end}<\bar{\rho}_r^{\rm end}$}, as found by Ref.~\cite{Bettoni:2021zhq},
and using that \mbox{$\bar{\rho}_\psi^{\rm end}=\frac14\lambda\langle\psi^2\rangle^2$},
we find
\begin{equation}
  \langle\psi^2\rangle_{\rm end}
  <6\sqrt 2\,(\hat\xi/\sqrt\lambda)\,\bar{H}_{\rm end}^2 \, .
\label{psiendbound}
\end{equation}
where we considered Eq.~\eqref{rhorend}. 
Considering that the typical value of
the amplitude of the oscillating condensate at a given location is of the order
\mbox{$|\psi|\sim\sqrt{\langle\psi^2\rangle}$}, we can estimate how it evolves
after the end of inflation. Indeed, because
\mbox{$\frac14\lambda|\psi|^4=\bar{\rho}_\psi\propto \bar{a}^{-4}$}, we have
\mbox{$|\psi|\propto \bar{a}^{-1}$}. Thus, we obtain
\begin{eqnarray}
&& |\psi|_{\rm reh}=\frac{\bar{a}_{\rm end}}{\bar{a}_{\rm reh}}|\psi|_{\rm end}\nonumber\\
\Rightarrow &&
|\psi|_{\rm reh}<6\left(\frac{2}{\lambda}\right)^{1/4}\,\hat\xi^{3/2}\,\frac{\bar{H}_{\rm end}^2}{m_P}\,,
\label{psirehbound}
\end{eqnarray}
where we used Eqs.~(\ref{aratio}) and (\ref{psiendbound}).

The quadratic term in Eq.~(\ref{Vpsi}) takes over from the quartic term at
a critical value $\psi_{\rm x}^2$ when
\mbox{$\frac12 m^2\psi_{\rm x}^2=\frac14\lambda\psi_{\rm x}^4$}, which suggests
\begin{equation}
\psi_{\rm x}^2=2m^2/\lambda\,.
\label{psix}
\end{equation}
To make sure that this does not happen until reheating, we simply require
\mbox{$\psi_{\rm x}^2\leq|\psi|_{\rm reh}^2$} ($|\psi|$ is reducing in time).
Then the bound in Eq.~(\ref{psirehbound}) results in the bound
\begin{equation}
m^2<m_{\rm max}^2\equiv 18\sqrt{2\lambda}
\,\hat\xi^3\,\frac{\bar{H}_{\rm end}^4}{m_P^2}\,.
\label{mbound}
\end{equation}  
The above is too strict because, if \mbox{$\bar{\rho}_\psi\ll\bar{\rho}_r$} after the end of
inflation, then $\bar{\rho}_\psi$ can remain subdominant until reheating even if
the quadratic term in $V(\psi)$ takes over before reheating. So the above bound
is sufficient but not, strictly speaking, necessary. Its numerical value may be estimated using the range obtained in Eq.~(\ref{hatxirange}). Taking \mbox{$\lambda\sim 1$}, we find
\begin{equation}
 10^2\;{\rm GeV}\lesssim m_{\rm max} < 
 10^{15}
  \;{\rm GeV}.
\end{equation}
The lower bound in the above might be unrealistic because such a particle could have been already observed in the LHC. But we see that the mass range extends well above the TeV scale so there is no real conflict with the observational data. 

\section{Energy density of gravitational radiation at the end of inflation}
\label{rhoGW}

The energy density of gravitational waves is 
\cite{Dufaux:2007pt}
\begin{equation}
    \rho_{_{\rm GW}}(\tau,{\bf x})=\frac{\langle h'_{ij}(\tau,{\bf x})h'_{ij}(\tau,{\bf x})\rangle}{32\pi G a^2}\,,
\end{equation}
where the prime denotes derivatives with respect to conformal time $\tau$, $h_{ij}$ are the spatial components of the metric perturbation and we consider superhorizon scales. We consider the Einstein frame and omit the overbar for simplicity. Switching to momentum space, we can define the density parameter of gravitational waves per logarithmic momentum interval
\begin{equation}
    \Omega_{_{\rm GW}}(\tau, k)\equiv\frac{1}{\rho_c}\frac{{\rm d}\rho_{_{\rm GW}}(\tau,k)}{{\rm d}\ln k}\,,
\end{equation}
where \mbox{$\rho_c=3H^2m_P^2$}.
Gravitational waves generated by inflation obtain a predominantly scale-invariant superhorizon spectrum given by \cite{Caprini:2018mtu}
\begin{equation}
    \Delta_h^2(k)=\frac{2}{\pi^2}
    \left(\frac{H_{\rm end}}{m_P}\right)^2\left(\frac{k}{k_*}\right)^{n_t}
    \simeq\frac{2}{\pi^2}
    \left(\frac{H_{\rm end}}{m_P}\right)^2\,,
    \label{Delta}
\end{equation}
where \mbox{$|n_t|\ll 1$} ia the tensor spectral index, the star denotes the pivot scale and
\begin{equation}
    \langle h_{ij}(\tau,{\bf x})
    h_{ij}(\tau,{\bf x})\rangle=\int\frac{{\rm d}k}{k}\Delta_h^2(k)(\tau,k)\,.
\end{equation}
Then, the final expression of the stochastic gravitational wave background from inflation is \cite{Boyle:2005se}
\begin{equation}
    \Omega_{_{\rm GW}}(\tau,k)=\left(\frac{a_k}{a(\tau)}\right)^4\left(\frac{H_k}{H(\tau)}\right)^2\frac{\Delta_h^2(k)}{24}\,,
\end{equation}
for arbitraty evolution $a(\tau)$, where \mbox{$a_kH_k=k$} coresponds to horizon re-entry of scale $k$. Evaluating the above at the end of inflation and integrating over all superhorizon modes we obtain
\begin{equation}
    \Omega_{_{\rm GW}}^{\rm end}\simeq\frac{1}{12\pi^2}\left(\frac{H_{\rm end}}{m_P}\right)^2,
\end{equation}
where we considered that the integral is dominated by the highest $k$ (i.e. $k_{\rm end}$). Using that \mbox{$\rho_{_{\rm GW}}=\Omega_{_{\rm GW}}\rho_c$}, we find
\begin{equation}
    \rho_{_{\rm GW}}\simeq\frac{1}{4\pi^2}H_{\rm end}^4,
\end{equation}
which agrees nicely with the estimate in Ref.~\cite{Ford:1986sy}.

\bibliography{Pala_Quint}

\end{document}